\documentclass[a4paper,11pt]{article}

\usepackage{jinstpub} 
\usepackage{placeins}
\usepackage{graphicx}
\usepackage{siunitx}

\usepackage{amsmath}

\usepackage[toc,page]{appendix}
\usepackage{comment}
\usepackage{lineno}

\usepackage[export]{adjustbox}

\usepackage{todonotes}

\title{ High Voltage Delivery and Distribution for the NEXT-100 Time Projection Chamber}
\author[1,a]{C.~Adams\note[a]{Now at NVIDIA.},}
\author[2]{H.~Almaz\'an,}
\author[3]{V.~\'Alvarez,}
\author[1]{K.~Bailey,}
\author[4]{R.~Guenette,}
\author[5]{B.J.P.~Jones,}
\author[1]{S.~Johnston,}
\author[5]{K.~Mistry,}
\author[2,6]{F.~Monrabal,}
\author[5]{D.R.~Nygren,}
% ...
\affiliation[1]{
Argonne National Laboratory, Argonne, IL 60439, USA}
\affiliation[2]{
Donostia International Physics Center, BERC Basque Excellence Research Centre, Manuel de Lardizabal 4, San Sebasti\'an / Donostia, E-20018, Spain}
\affiliation[3]{
Instituto de Instrumentaci\'on para Imagen Molecular (I3M), Centro Mixto CSIC - Universitat Polit\`ecnica de Val\`encia, Camino de Vera s/n, Valencia, E-46022, Spain}
\affiliation[4]{
Department of Physics and Astronomy, Manchester University, Manchester. M13 9PL, United Kingdom}
\affiliation[5]{
Department of Physics, University of Texas at Arlington, Arlington, TX 76019, USA}
\affiliation[6]{
Ikerbasque (Basque Foundation for Science), Bilbao, E-48009, Spain}
% ...

\emailAdd{leslie.rogers776@gmail.com}

\abstract{A critical element in the realization of large liquid and gas time projection chambers (TPCs) is the delivery and distribution of high voltages into and around the detector.  Such experiments require of order tens of kilovolts to enable electron drift over meter-scale distances.  This paper describes the design and operation of the cathode feedthrough and high voltage distribution through the field cage of the NEXT-100 experiment, an underground TPC that will search for neutrinoless double beta decay $0\nu\beta\beta$.  The feedthrough has been demonstrated to hold pressures up to 20~bar and sustain voltages as high as -65~kV.  The TPC is operating stably at its design high voltages. The system has been realized within the constraints of a stringent radiopurity budget and is now being used to execute a suite of sensitive double beta decay analyses.}

\begin{document}

\maketitle

\flushbottom

\section{Introduction }
\label{sec:intro}

\begin{figure}
    \centering
    \includegraphics[width=.75\textwidth]{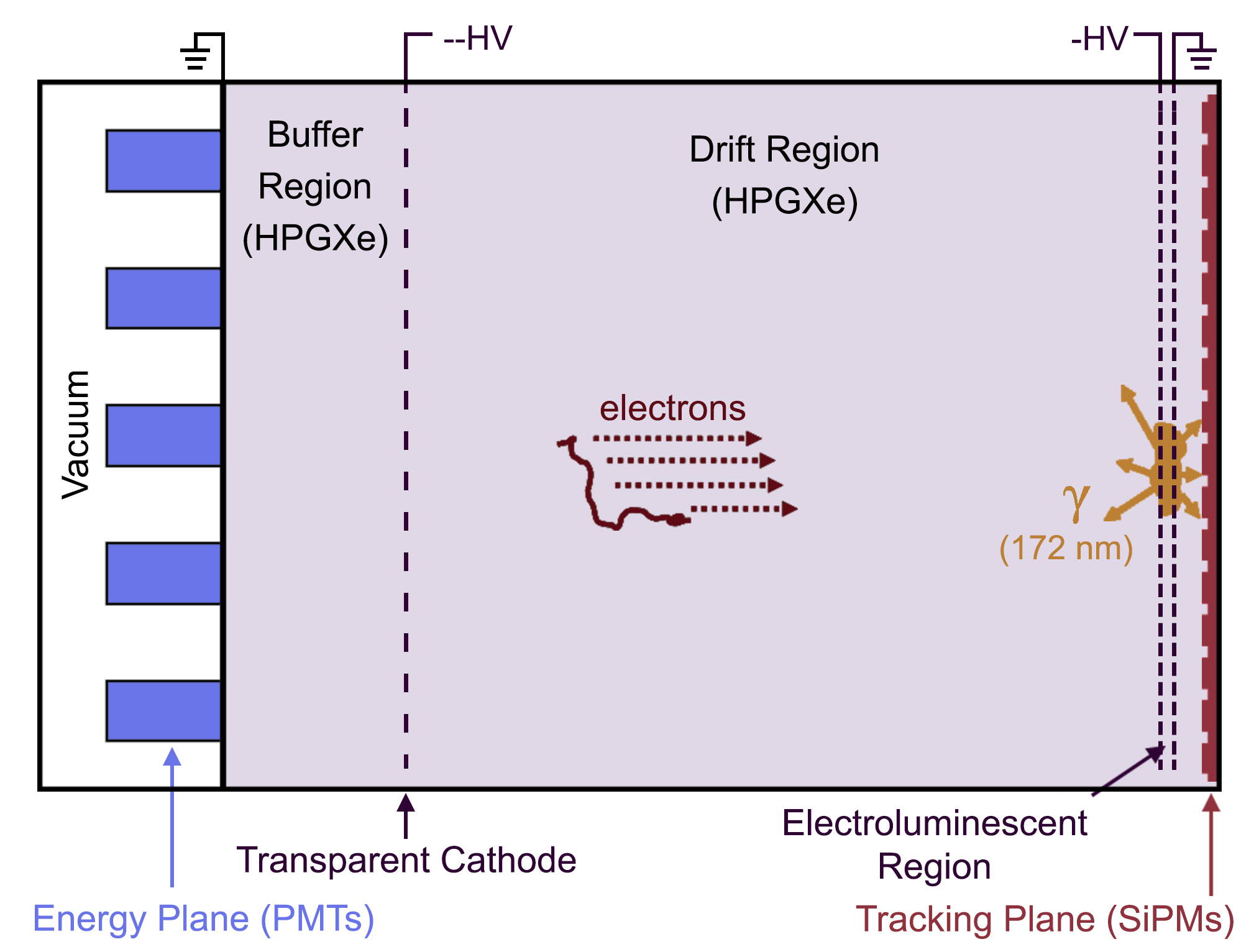}
    \caption{Diagram of the NEXT-100 TPC which is held in pressurized xenon (HPGXe) with the energy readout plane at the left consisting of 60 photomultiplier tubes (PMTs) held in vacuum. The cathode is held at a high negative voltage with a buffer region behind it to safely step the voltage up from ground. The drift field is a well controlled moderate electric field that moves electrons over to the electroluminescence region where there is a high electric field that creates light amplification via electroluminescence, collected via the PMTs for an energy readout, and 3584 silicon photo multipliers (SiPMs) for the tracking readout.}
    \label{fig:NEXT100Outline}
\end{figure}

\begin{figure}
    \centering
    \includegraphics[width=.95\textwidth]{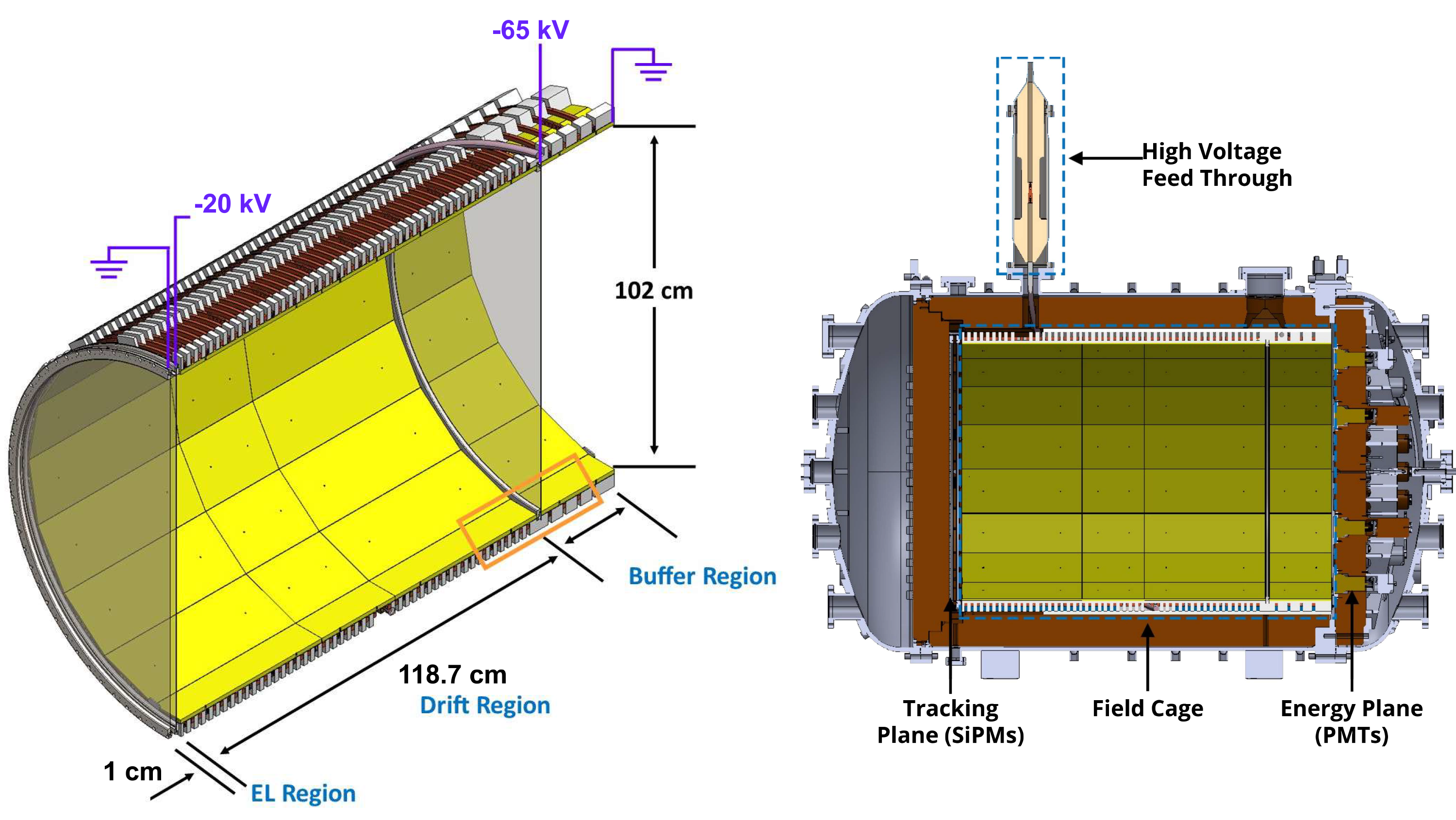}
    \caption{Left: Sketch of the field cage and TPC system, with the relevant high voltages indicated. See text for details and Fig.~\ref{fig:FC} for a close-up of the components contained in the orange rectangle. Right: Cross Section of the NEXT-100 detector design showing the field cage within a pressure vessel and connected to a high voltage feedthrough which sits on top and is explained in more detail in Section \ref{sec:Design}.}
    \label{fig:NEXT100FC}
\end{figure}

The Neutrino Experiment with a Xenon TPC (NEXT) program is the product of an international collaboration aiming to develop a series of high pressure gas xenon Time Projection Chambers (TPCs) to search for evidence of the Majorana nature of the neutrino through observation of neutrinoless double beta decay ($0\nu\beta\beta$).  The progression of NEXT detectors has increased exponentially in scale between generations and aims to culminate in ton- to multi-ton scale modules that will execute a world-leading search for $0\nu\beta\beta$. Small to mid-scale demonstrators have included NEXT-DBDM~\cite{alvarez2013near} NEXT-DEMO~\cite{Alvarez2013DEMO} and NEXT-White~\cite{WHITE_TDR}. These experiments have proven the detector technology capable of excellent energy resolution with 1\% full-width half maximum (FWHM) at 2.5 MeV~\cite{NEXT:2019qbo}, along with performing a measurement of the two-neutrino double beta decay background in $^{136}$Xe~\cite{NEXT:2021dqj}, and a first search for $0\nu\beta\beta$~\cite{NEXT:2023jsn}. The current phase of the experiment is called NEXT-100~\cite{Alvarez2012NEXT100TDR} and schematically shown in Figs. \ref{fig:NEXT100Outline} and \ref{fig:NEXT100FC}.  

Noble element TPCs operate in a variety of phases, both liquid and gas, for numerous types of physics searches today~\cite{universe5030073,t2knd280tpccollaboration2010time, ACKERMANN2010141,She_2023,Ferrario:2015kta}. As these experiments grow in size, the drift distances become longer, and therefore, the voltages required increase. This scaling has often resulted in TPC experiments unsuccessfully reaching their design field goals~\cite{TomShuttKilotonne}. 

An advantage of gaseous xenon is that the track length of a decay particle is relatively long compared with condensed phases, around 20~cm in 13.5 bar. When a double beta decay occurs within NEXT-100, it creates an initial light signal from the excitations created when the decay particles deposit their energy in the xenon, as well as ionization electrons. These electrons can be imaged in three dimensions, and so the particle path can be reconstructed. A moderate electric field ($\cal{O}$300-400 V/cm), the drift field, draws the ionized electrons to an electroluminescence region. This region has a high electric field ($\cal{O}$10 kV/cm) where the electrons gain sufficient energy to excite the xenon atoms within their path. When an excited xenon atom de-excites it then releases a photon, resulting in an electroluminescent gain process whereby each electron can generate $10^2-10^3$ secondary scintillation photons as it crosses the electroluminescence (EL) region. The XY location of these photons is imaged via a tracking plane that consists of an array of silicon photomultipliers (SiPMs). The third dimension of the event is encoded in the time of which the electrons produce light in the EL region, which can then be converted to a Z component via the known electron drift velocity. The event energy is also recorded, both in the SiPM array, and with finer precision using a high photo-cathode coverage plane of photomultiplier tubes that count the total number of photons created.  The number of photons is proportional to the initial number of electrons with remarkably low fluctuations, after correcting for losses during drift and local detector response maps. These maps are continuously produced through calibration with a circulating gaseous $^{83\textnormal{m}}$Kr  source~\cite{martinez2018calibration}.  A measurement of energy with a resolution better than 1\% FWHM is enabled~\cite{renner2019energy} via collection of secondary scintillation from the EL region.  In this way, the NEXT EL TPCs represent precise calorimetric 3D-imaging detectors.  Operation of both the TPC and the EL region requires the supply and distribution of a substantial DC voltage, of order 70~kV.  A 3D model of the TPC system for NEXT-100 with the relevant high voltage connections indicated is shown in Fig.~\ref{fig:NEXT100FC}.

The drift field requirements are driven by a number of factors. When the ionization track is created, the electric field must be sufficient to prevent $e^-$+Xe$^+$ recombination.   The drift velocity should also be sufficiently high that the readout window associated with one event is short enough to make pileup unlikely, or to induce performance degradation of the energy resolution due to electron attachment on electronegative impurities in the gas. Uniformity of the field in the fiducial volume must be maintained in order to protect the topology of the initial track, which encodes the type of event that occurred and allows separation of signal from backgrounds~\cite{Ferrario:2015kta}.  Drifting electrons also undergo both transverse and longitudinal diffusion, and the diffusion scale is minimized for larger drift voltages and hence shorter drift times. 
 As such, higher drift fields lead to more faithful images of the original ionization deposits.  All of these factors drive a requirement for an electric field of order 300-400 V/cm, which is homogeneous at the sub-percent level. This field is created by a ``field cage'', in which conducting rings are held at graded potentials to ensure a uniform, parallel field in the drift volume. An overview of the field cage design for NEXT-100 can be found in sec.~\ref{Sec:FieldCage}.

To generate the drift field at design specification in NEXT-100, 400~V/cm in 13.5~bar of xenon \cite{next100CDR}, a high voltage is applied to the cathode and stepped down along the field cage using a resistive divider inside the vessel. This voltage, generated by a DC power supply outside the pressure vessel, must enter through a penetration called the ``cathode high voltage feedthrough'' in order to meet the cathode.  This component is the subject of sec.~\ref{sec:Design}.  The design specification for the cathode voltage is -65~kV, which is sufficient to both enable the required drift field and the needed voltage step up of the electroluminescent region.  The cathode itself is formed from a stretched, photochemically etched mesh of 127~\si{\micro\meter} wire size over a two-part silicon bronze frame.  Its design, as well as that of the EL region, is described in detail in \cite{Mistry_2024}. 

NEXT-100 also includes two further high voltage feedthroughs, which provide voltages for the EL region. Since the operating parameters of this sub-system are similar to the prior iteration,  NEXT-White,  the previous EL region feedthroughs were re-used for these parts~\cite{WHITE_TDR}. Supply and distribution of high voltage to the cathode and field cage in NEXT-100, however, required significantly more advanced components than used in previous NEXT experiments \cite{Alvarez2013DEMO,WHITE_TDR}. Realizing these involved a program of design, R\&D, and characterization, as well as adherence to a strict radiopurity budget, described in sec.~\ref{sec:radiopurity}.  This  paper focuses on development and installation of the cathode HVFT and The field cage for the NEXT-100 experiment.   These components have been instrumental in enabling the rapid commissioning and operation of NEXT-100, which is reached and held its design high voltages following a brief commissioning period in argon and xenon gases.  Validation, commissioning and performance of the system are discussed in sec.~\ref{sec:validation}.  

\section{Field cage design}
\label{Sec:FieldCage}

\begin{figure}[t]
    \centering
    \includegraphics[width=.9\textwidth]{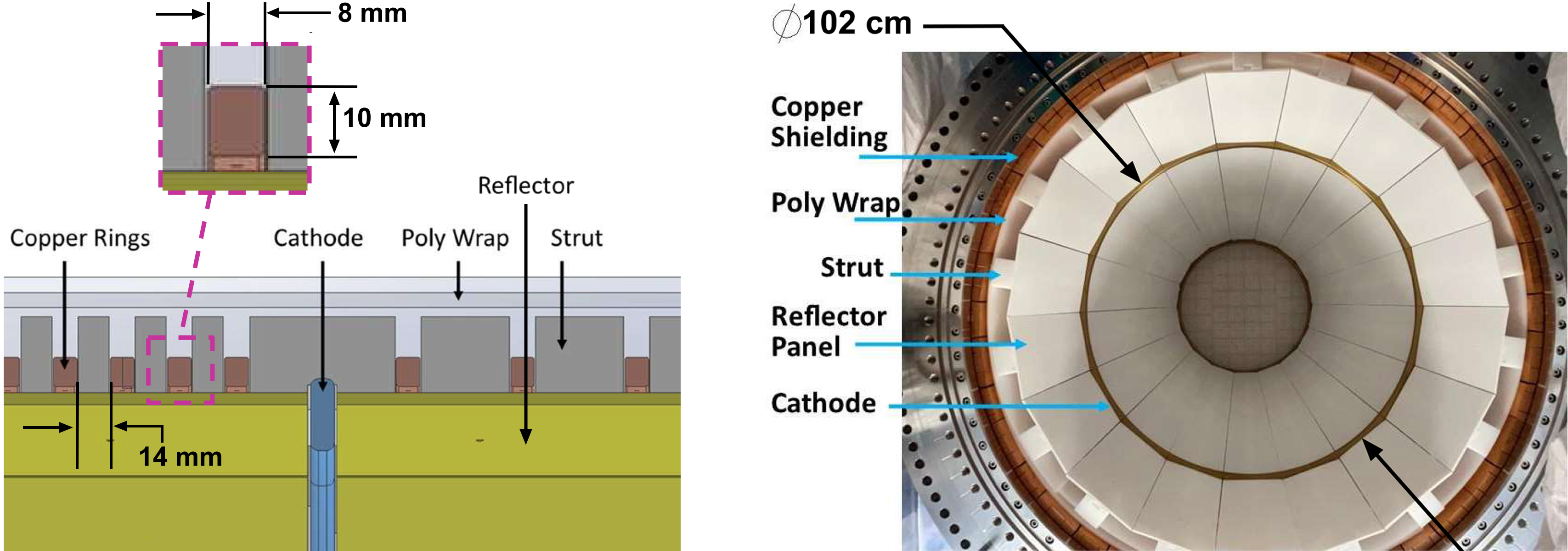}
    \caption{Left: Labeled cross section of the field cage centered at the cathode with the buffer region on the right and drift to the left. Dimensions are provided for the cross section of the copper rings. Right: Internal view of NEXT-100 from the energy plane side looking towards the SiPMs of the tracking plane. The gold ring is the edge of the cathode which is holding a 95\% transparent mesh. Note that the copper rings and resistor boards are not visible from this angle.}
    \label{fig:FC}
\end{figure}

\begin{figure}[t]
    \centering
    \includegraphics[width=.99\textwidth,valign=c]{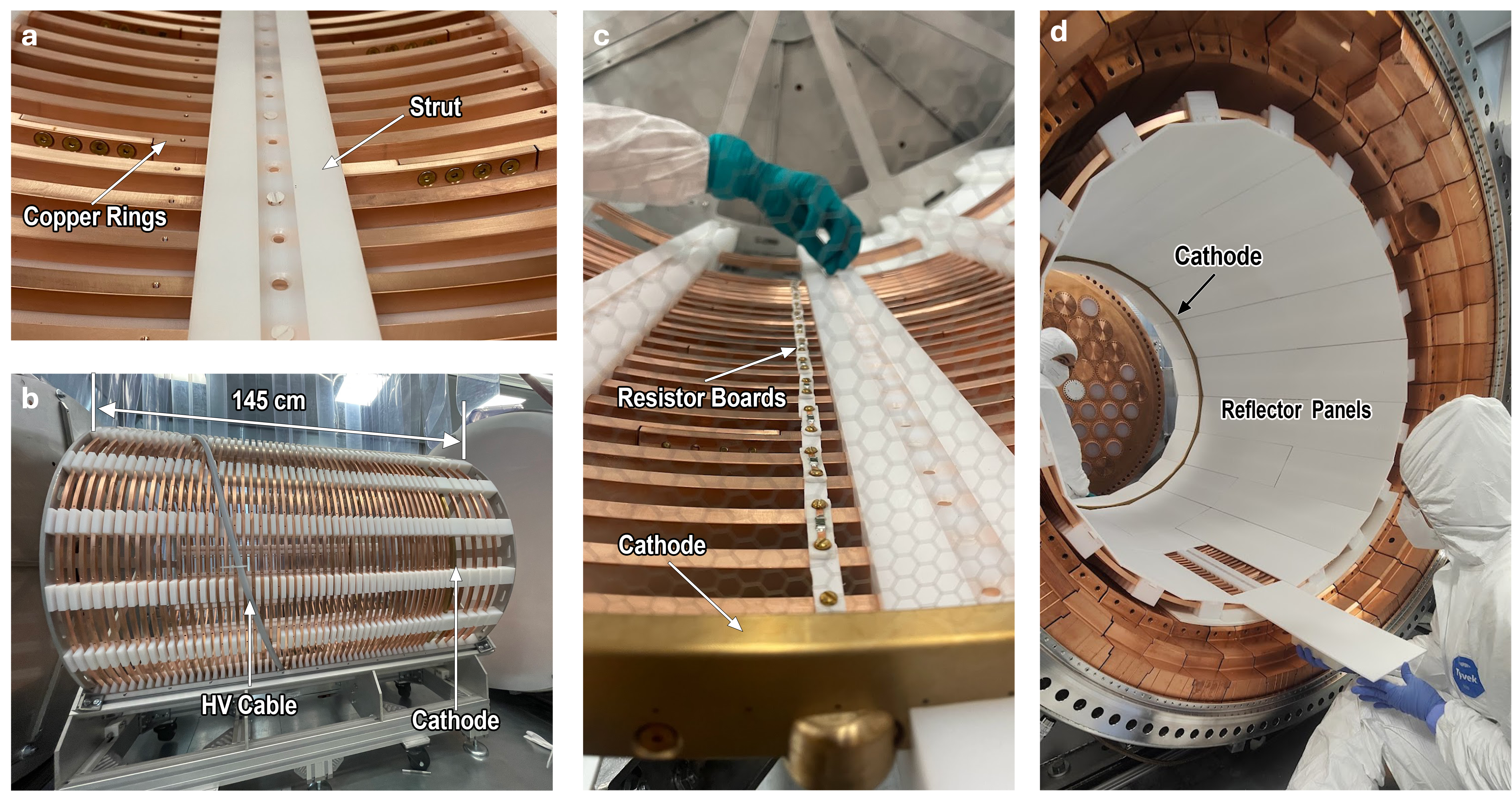}

    \caption{Details of the field cage construction. a. connections of the segmented copper rings; b. assembled cage structure with all struts and rings installed and HV cable visible around the outer circumference; c. one of three parallel resistor chains running up the inside of the field cage; d. installation of the wavelength shifting PTFE reflectors. See text for details.}
    \label{fig:FCDetails}
\end{figure}

The field cage consists of 52 copper rings (48 in the drift region and 4 for the buffer region) supported on 18 high density polyethylene (HDPE) struts, as shown schematically in Fig.~\ref{fig:FC}, and in detail in Fig.~\ref{fig:FCDetails}. Attached to the struts are polytetrafluoroethylene (PTFE) reflectors. These components are wrapped in a layer of HDPE ``Poly Wrap" to insulate the high voltage electrodes from the internal copper shielding that absorbs gamma radiation produced in the stainless steel of the vessel walls.

The complete field cage is 142.7~cm in length and 110.5~cm in diameter of which 118.7~cm comprises the drift region (on the EL-region side of the cathode). The field cage design was optimized to maximize the detector fiducial volume within the boundaries of the inner copper shielding, as well as to minimize the mass of structural materials for radiopurity purposes. To achieve both goals, the design is made with interconnecting components that slot together during assembly, requiring minimal passive scaffolding and fixtures, and making efficient use of the available volume. The entire field cage was assembled outside the pressure vessel as free-standing horizontal cylinder and was slid inside as one piece. A schematic of the field cage can be seen in Figure~\ref{fig:FC}.

The copper field-shaping rings grade the potential uniformly along the drift field, stepping down the voltage applied at the cathode. The rings have an internal (external) diameter of 101.4~cm (103.8~cm), a shape of a rounded rectangle of dimensions 10~mm thickness, 8~mm width, 1~mm rounded corners, and cross sectional area 1.2~cm$^2$. They are made in three segments and joined by 12 brass countersunk screws in order to facilitate machinability (Fig.~\ref{fig:FCDetails},a). Finite element analyses were performed to validate that the expected sag from using joined rings was not prohibitive to maintaining the structural form of the assembled field cage.  The copper rings ultimately provide the main structural elements that maintain the shape of the cylinder, with the HDPE staves serving primarily to fix each ring relative to the others.  The edges of the copper rings are rounded to avoid high electric field concentrations that could lead to dielectric breakdown or corona in the surrounding xenon gas. 

Between each two rings is a 1.4~cm edge-to-edge gap, spanned by three parallel arrays of film-chip-film resistors (CHRV100MEDKR-ND from Vishay Techno) of resistance 100~M$\Omega$. The use of parallel resistor chains provides redundancy in the event of a resistor failure, and positioning inside the cage avoids exposure of the resistor edges to large electric fields. The resistors are soldered to small mounting boards of dimension 28~mm $\times$ 6~mm and 2.7~mm thickness manufactured from Cuflon (a board with a core of PTFE and planes of copper) from the Polyflon company. One of the resistor board chains can be seen in Fig.~\ref{fig:FCDetails}, c, where resistor boards are installed upside down every other ring to facilitate their installation. The boards are mounted on the inside of the field cage rings to minimize electrical discharges to the inner copper shield from sharp surfaces on the resistors and each board is mounted 120 degrees from the other parallel boards between each field cage ring gap. Both board material and resistors were screened for radiopurity~\cite{alvarez2013radiopurity} and found to be comfortably within acceptable limits. A special resistor board spans the 2.45~cm gap between the last field cage ring and the cathode with two resistors in parallel and one in series replicated three times around the azimuth to provide appropriate field grading over the slightly large field-cage-to-cathode gap. The total electrical resistance from the cathode to the last ring of the TPC, is designed to be 1.65~G$\Omega$ and was measured to be 1.68~G$\Omega$ during installation using a series picoammeter. This small difference has no important implication for the operating drift field. 

In the opposite axial direction from the drift volume, a shorter series of four rings with 3.8~cm gap between each ring extends from the cathode toward the energy plane, which is held at ground. The field uniformity requirements in this region are far less stringent, so in order to minimize both radiopurity load and failure modes these rings do not have resistors between them, instead relying on the resistivity of the polyethylene frame to grade the voltage.

The HDPE struts provide structural support for the field cage and run the entire detector length. To attach to the copper rings, they have ``comb'' features cut into them. Similarly, the HDPE struts have a groove on the inside to hug the cathode and, as such, were assembled piece by piece around the cathode. This scheme is shown in Figure~\ref{fig:FC}. The copper rings are attached with nylon screws to the HDPE struts, and one of these screws, as can be seen in Fig.~\ref{fig:FCDetails}, a.

Attached to the field cage HDPE struts are individual panels of PTFE reflector material with 0.5~cm thickness. These reflectors slide into the struts with a dovetail joint to create a light-tube, lining the active volume to enhance photon collection efficiency. These panels are also coated in the wavelength shifter Tetraphenyl Butadiene (TPB) with a thickness of a few microns. Although the sapphire windows of NEXT-100 will transmit UV light, and the photomultiplier tubes have some sensitivity the xenon emission wavelength of 175~nm, the better PMT quantum efficiency at the TPB emission peak leads to improved energy resolution with the application of a wavelength shifter.  A program of R\&D has quantified the reflectivity of NEXT-100 style PTFE panels vs thickness \cite{haefner2023reflectance}, and it was found that 0.5 cm thick panels offer excellent reflectivity while avoiding unnecessary additional material mass.  It is notable that this is somewhat different to the optimal thickness for liquid xenon experiments~\cite{neves2017measurement, haefner2017reflectance}, where PTFE reflectivity is significantly enhanced due to the high refractive index of the liquid leading to increased fresnel reflectivity~\cite{kravitz2020measurements}, which is a thickness-independent effect. In the gas phase, the reflectivity is primarily due to diffusion of the photons within the PTFE volume, and it therefore exhibits a more significant thickness dependence.  

Where previous NEXT detectors have used a solid light tube, these panels were produced in sections to make coating with TPB more straightforward and controllable, and installation more manageable. Installation of one of the panels is shown in Fig.~\ref{fig:FCDetails}, d. Due to their modularity, it is anticipated that a similar design can be used for larger future detectors, following validation of performance in NEXT-100. 

\begin{figure}[t]
    \centering
        \includegraphics[height=.35\textwidth]{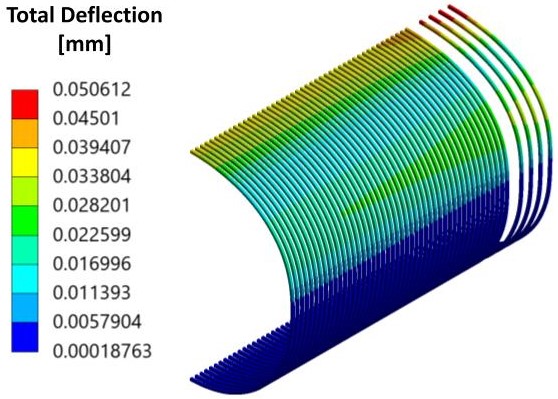}
    \includegraphics[height=.35\textwidth]{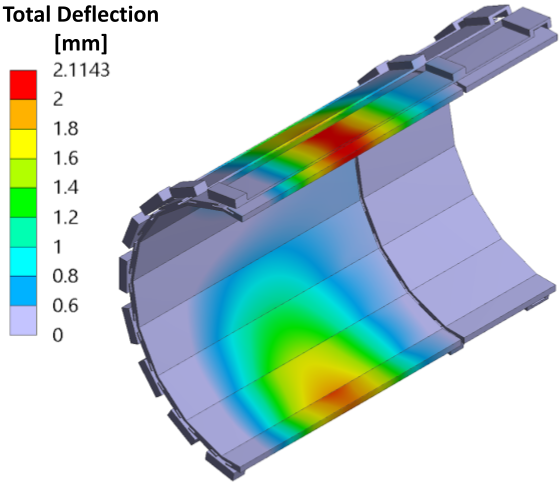}
    \caption{Left: Total deformation expected from the copper rings in the field cage which produce the mechanical support to hold everything together. This simulation took into account the weight of all the plastics and the rings themselves with the largest deformation occurring at the top of the buffer region where there is a larger ring spacing resulting in the rings each having to support more mass. Even in this area there is trivial deformation expected of 0.05~mm. Right: Total deformation expected from the field cage HDPE struts and light panels once assembled. The deformation simulation took into account the weight of the plastics and demonstrates a robust design with a maximum of 2.1~mm  near the center of the drift field region.   }
    \label{fig:FCdeflect}
\end{figure}

The mass of the field cage is 270.7 kg (excluding the Poly Wrap) and is overwhelmingly dominated by copper, HDPE, and PTFE, at 177.9~kg, 54~kg, and 38.8kg, respectively. All of this weight is supported via the copper rings. To check if the rings could support this structure, a finite element analysis was used to evaluate the rings throughout the system as shown in Figure~\ref{fig:FCdeflect} Left. It was determined that the structure was mechanically robust with inconsequential deformation in the rings, with less than 0.05~mm of deflection from the nominal cylindrical shape everywhere within the region of interest. A stress analysis was also performed for the rings and found to be trivial throughout, with a max of 2.71~MPa at the peak deflection area. 

Since the PTFE panels are a significant portion of the weight and cantilever out on either side of the dove tail where it slides into the HDPE struts, this deformation mode was also studied. The results can be seen in Figure~\ref{fig:FCdeflect} Right. There is more deformation than the copper rings alone, with a maximum of 2.1~mm  sag in the center area of the drift region. This was determined to be within our tolerances, and is not expected to impact either the high voltage distribution or the uniformity of light collection. 

The Poly Wrap is made from two layers of 0.5~cm thick of HDPE. It is wrapped around the field cage to insulate the high voltage electrodes from the internal copper shielding. The inner layer makes a nearly complete turn of the cylinder leaving just a thin seam at the bottom of the detector. The outer layer runs around 40\% the circumference to cover this remaining seam. This structure prevents the occurrence of an electrical breakdown from the cathode to the pressure vessel across the few-cm xenon gas gap that separates them.  The dielectric strength of HDPE sheet was tested in parallel plate conditions prior to installation and proven to exceed the design strength required at the cathode with a safety factor of at least two.

A high voltage cable is connected to the cathode and wound around the field cage (as can be seen in Fig.~\ref{fig:FCDetails},b) until it gets near the EL region, where it then exits from the top of the pressure vessel. Because the absolute voltages become smaller near the EL region, it becomes admissible to break the Poly Wrap near to this end of the field cage without risk of discharges to the copper shielding. The cathode high voltage cable thus emerges from the Poly Wrap near the EL region and turns outward to meet the cathode high voltage feed through. Along the path where the high voltage cable curves around the outside of the field the HDPE struts have shorter ridges to accommodate the cable outside the field cage but inside the Poly Wrap.

\begin{figure}[htb]
\centering
\includegraphics[width=.95\textwidth]{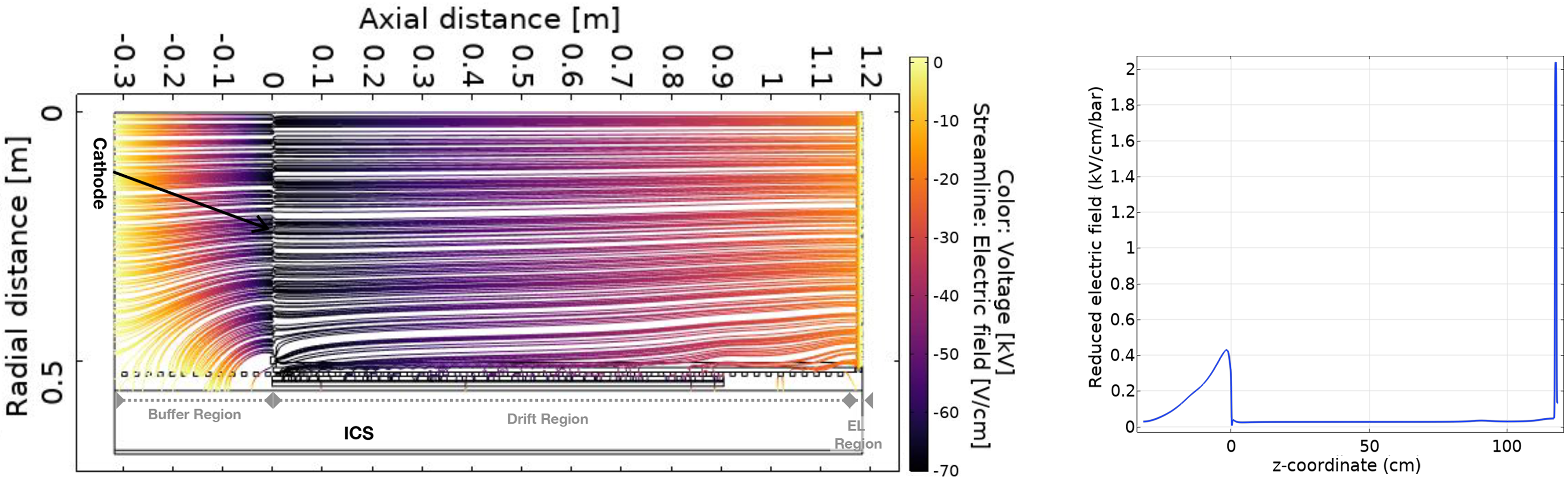}
\caption{Left: Simulations of the electric field in the NEXT-100 field cage using COMSOL assuming operating voltages of -70~kV at the cathode and -20~kV at the gate. This corresponds to a constant and homogeneous electric field of $\sim$\qty{400}{\volt\per\centi\meter} within the operating region of the detector. Right: Reduced electric field along z at a radial distance 20~mm from the PTFE walls. Both images are reproduced from \cite{nextcollaboration2025next100detector}. }
\label{fig:Voltage}
\end{figure}

Simulations of the electric field in the NEXT-100 field cage were performed using COMSOL and can be seen in Fig.~\ref{fig:Voltage}. These simulations use a 2D-axisymmetric approximation to evaluate the radial and axial components of the electric field. A 20~mm fiducial cut from the PTFE walls is found to be sufficient to ensure a uniform drift field for double beta decay measurements.

\section{Cathode high voltage feedthrough}
\label{sec:Design}

The centerpiece of the NEXT-100 high voltage system is the cathode high voltage feedthrough.  The HVFT has to be capable of stably holding voltages to the operating specification of -65~kV without charge-up, breakdown, or dielectric deterioration.  It delivers voltage from outside the vessel, i.e. air at atmospheric pressure, to the cathode which is mounted inside the inner copper shield, Poly Wrap, and field cage.  The required operational pressure conditions vary from vacuum during the pump-down phases to pressures of to 13.5 bar (the maximum certified pressure of the NEXT-100 vessel) of xenon gas during physics data taking. The feedthrough must be hermetic and provide an excellent seal even when pressurized, to avoid both loss of valuable $^{136}$Xe and inward diffusion of electronegative atmospheric impurities that impede TPC function.  In addition to these basic requirements, the feedthrough must be made from radiopure materials so as not to be a major source of backgrounds for the experiment.

\begin{figure}[t]
    \centering
    \includegraphics[width=.85\textwidth]{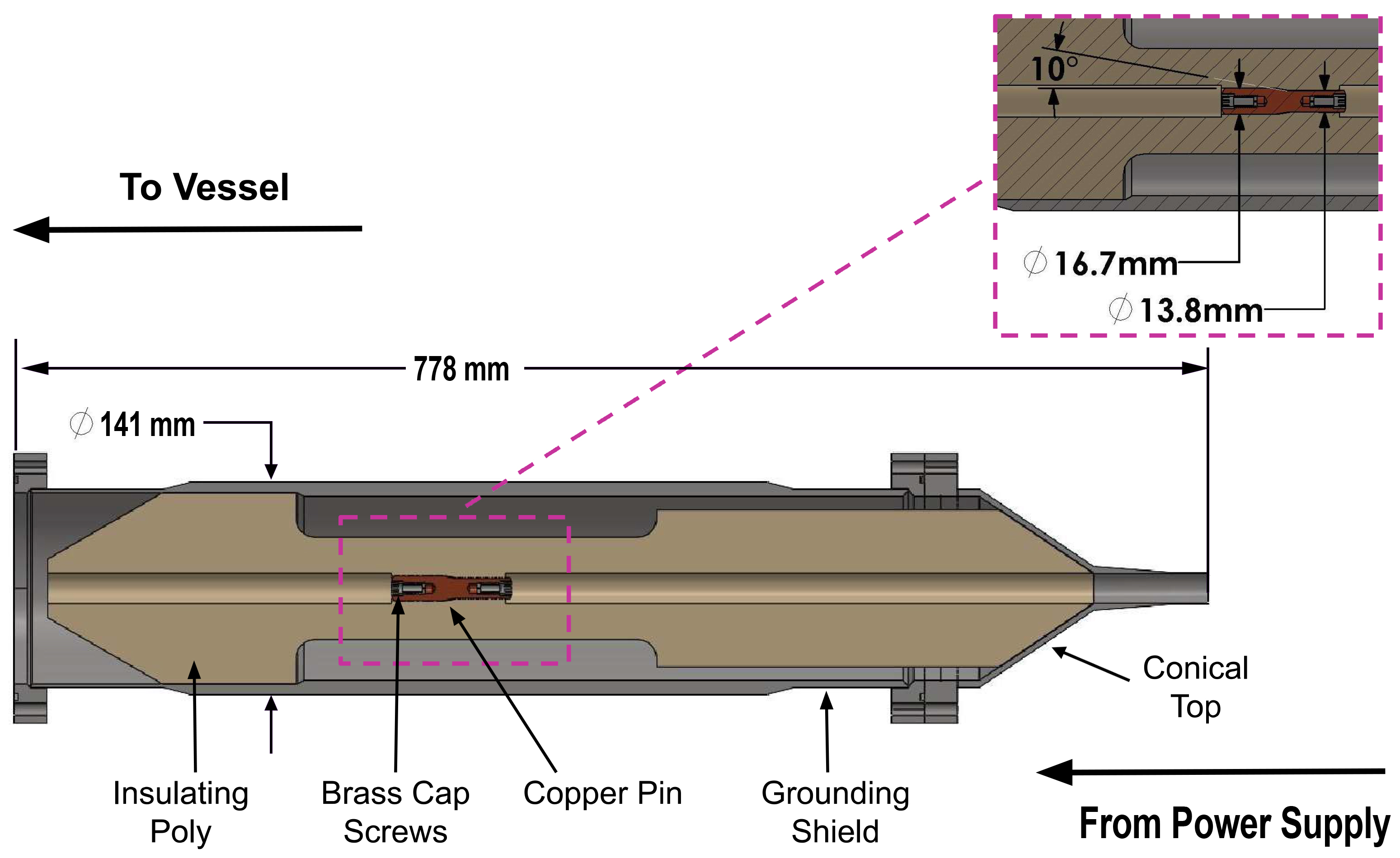}
    \caption{Cross section of the feedthrough showing the insulating polyethylene surrounding the copper pin, all within a grounding shield. The orientation shown has the power supply output coming in from the right and the vessel connection at the left. In the upper right, there is a close-up of the copper pin showing the difference in diameters and the taper between them. }
    \label{fig:HVFTcross}
\end{figure}

The HVFT is manufactured from four custom components, as shown in Fig. \ref{fig:HVFTcross}. The outer tube is a stainless steel 316 cylinder that acts as a grounding shield when connected to the grounded pressure vessel. Within the grounding shield, there is a thick insulating sheath made of Ultra High Molecular Weight (UHMW) polyethylene. Inside the sheath, the electrical connections between internal and external cables are securely made and protected from sparking to ground. The final critical component is the inner copper pin, which is precision-machined to cryo-fit inside the sheath and couple securely to both internal and external insulated high-voltage cables. 
 Electrical connections are made by feeding cables from either side of the HVFT until the ends are securely inserted into the copper connector pin. Once the banana plug connections are secure, the external part of the cable is clamped to the feedthrough with some compressive force applied to the inner cable insulator to keep the connection in place.

\subsection{Feedthrough design}

Although the HVFT necessarily contains large masses of material to avoid breakdown, it is situated mostly outside of the pressure vessel and the inner copper shielding,  so radiopurity constraints are relaxed relative to other NEXT field cage and voltage distribution components. Nevertheless, to avoid large backgrounds from materials near the detector and inside its shielding castle, materials were selected and sourced for their radiopurity, and material mass was minimized as far as reasonably possible.  To avoid the significant addition of sealing hardware, glues, or greases, interference fits were employed throughout. The interference fits are made by placing a cooled piece that would be oversized at room temperature within a warmer, undersized part and using pressure at the contact points created by thermal expansion to hold the components together. This fit must be robust enough to create a vacuum and pressure seal. This requires both radial and frictional forces,  creating stresses within materials and has to be carefully optimized to prevent fractures. The radial force induces radial stress, causing compression, tangential stress, and tension. In addition, the friction force induces shear stresses leading to both tension and compression in the system \cite{MechEngDesign}.
\begin{figure}[t]
    \centering
    \includegraphics[width=.85\textwidth]{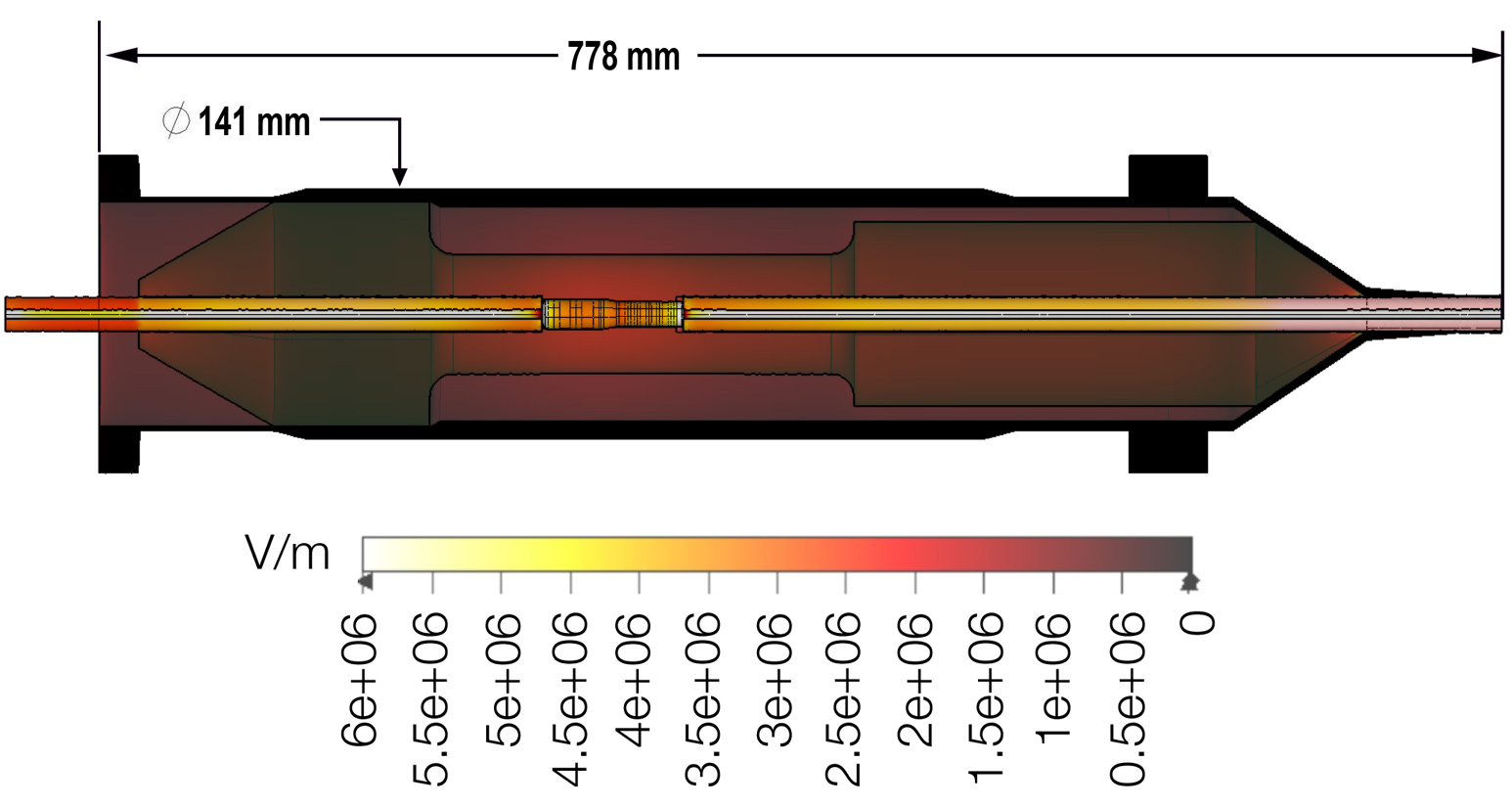}
    \caption{Simulation of the electric fields within the feedthrough when operating at 100~kV. The peak fields on the exterior of the polyethylene are less than 3~MV/m  while the fields within the cable channel, and especially near the copper pin, are as high as 10~MV/m in air.}
    \label{fig:ESim}
\end{figure}

The radii of the components for cryofit must thus be calculated and precisely machined since an over-tight fit will result in fractures, creating routes for leaks, whereas a loose fit will not produce a pressure seal.  This mechanical challenge is exacerbated by the difficulty of precisely machining large polyethylene parts. Several prototypes were developed at the Argonne National Laboratory machine shop in order to perfect production for effective cryofit. The results of this R\&D included the development of a custom tooling bit to machine the two diameters and the taper into the polyethylene in one step.  Other multi-step machining trials were explored, though the one presented here led to the highest quality surface finish.

The initial choice for an insulating sheath was HDPE rather than UHMW as it is the most stable while sustaining high electric fields \cite{HVinsulators}. However, our R\&D showed that when subjected to liquid nitrogen temperatures, it became brittle, leading to frequent cracks and leaks.  UHMW was then selected as a second material choice.  The coefficient of thermal expansion for UHMW polyethylene was determined empirically after the numbers in literature were found to be widely scattered \cite{plasticsmaterials,MACUVELE20171248} and ultimately unreliable compared to our in-house studies. This proved to be a crucial step for achieving a good seal. Each batch of UHMW was measured before machining in order to calculate the appropriate oversizing dimensions to fit inside the grounding tube. This was accomplished by measuring the width of a block of material under various temperatures to find the rate of expansion versus temperature and sizing the cryo-fit part accordingly.

The shape of the insulating polyethylene part of the HVFT was chosen to minimize high electric field concentrations that can seed breakdown or dielectric degradation. This is shown in Fig. \ref{fig:ESim}. The ends of the insulating layer are tapered to gradually reduce the field gradients and avoid creepage current or surface charge migration. Around the connector pin the outer diameter of the insulator is reduced, creating an air gap to force a reduction in the electric field around the copper pin where the non-uniform surface geometry creates varying fields.  The dielectric strength of the batch of UHMW used was 90~MV/m (per the manufacturers' specification), and our simulations showed that fields of this scale within the HVFT sheath were in principle possible, if any sharp edges remained around the connections. As an additional precaution, a VCR port was included on the first HVFT manufactured so that the gap around the polyethylene insulator could be filled with gases that have a higher dielectric strength than air, such as SF$_6$, or with vacuum grease, if needed. No breakdowns were observed during testing, discussed in more detail in Section \ref{sec:validation}, so the gas hookup was omitted from later versions.  Eventually, the final HVFT as installed on NEXT-100 was potted with a small quantity of epoxy in the upper cable space to minimize diffusion of xenon through the outer interference fit, in anticipation of the enriched $^{136}$Xe run.

The HVFT connects the cable from inside the vessel to the outer cable via the internal copper pin. This pin is machined from a rod of oxygen-free high thermal conductivity copper, and has two diameters with a 10 degree taper between them as shown in Fig. \ref{fig:HVFTcross}. The two steps and taper are larger on the vessel end as an added safety measure to prevent the connector pin from becoming a projectile under failure at high pressure. The copper connector has two base diameters on which several trapezoidal protrusions were machined. These protrusions were intended to increase interference with the polyethylene and keep any superficial defects on the surface from compromising a complete pressure seal. Each end of the connector pin is tapped and counter bored to accept a brass 5/16” socket head cap screw (SHCS) that was modified to accept a short banana plug. Once inserted into the SHCS and connector pin, the cable insulator sits flush against the connector pin, avoiding space for dielectric discharges.

The holes in the HDPE sheath of the HVFT where the cables are inserted are machined to a diameter larger than that of the cable with the grounding braid removed. These holes should be small enough to direct the cable straight into the pin, but (in particular on the inside) large enough not to trap gas to create a virtual leak when at vacuum or pressure. Furthermore, in early prototypes we found that if the fit was too tight, it would compress air behind the cable during insertion and pushes the cable back out of place, either immediately or over time.  The ideal diameter of the holes in the sheath was determined empirically based on these considerations to be {20.3~mm}, {1~mm} larger than the cable diameter.

\begin{figure}[t]
    \centering
    \includegraphics[width=\textwidth]{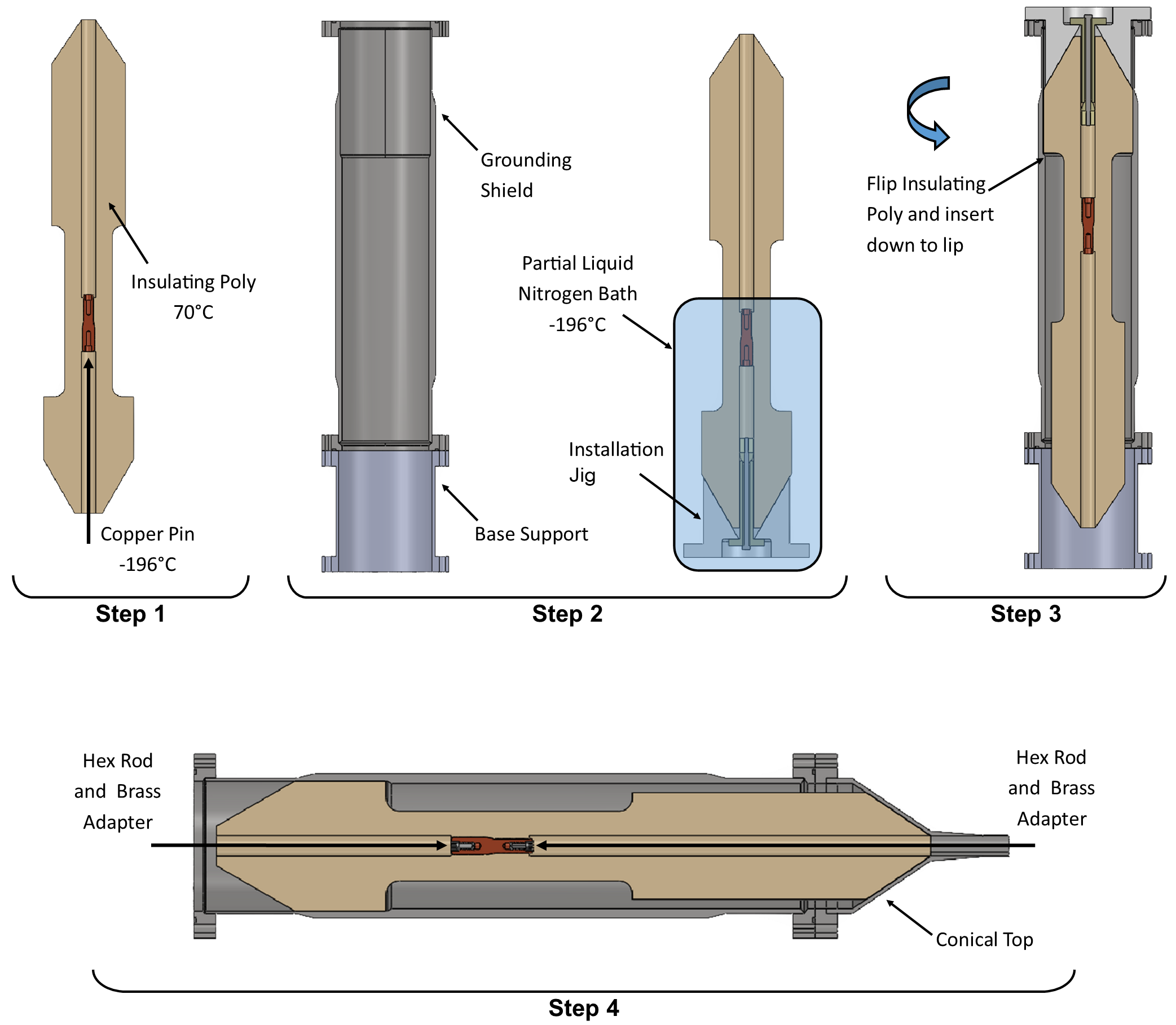}
    \caption{Assembly procedures for the HVFT. Step 1 is freezing the pin and heating the polyethylene to reduce and expand the diameters, respectively, before inserting them together for an interference fit. Step 2 mounts the grounding tube to a base while freezing the bottom of the insulating polyethylene.  In order for gravity to aid in the cryo-fit, in step 3 the part is flipped so that the cooled end is resting against the grounding tube and the non-cooled side extends freely into the base support. Step 4 adds adapters to the copper pin to be able to take banana plugs from the HV cables later as well as attaching the conical top.}
    \label{fig:HVFTassy}
\end{figure}

\subsection{Assembly procedure}
\label{Sec:HVFTassyProcedure}
The HVFT assembly requires minimal tooling, as most part matings are accomplished using interference fits. The only tool required was a 5/16" hex rod at least {50~cm} long and a threaded rod at least  $\sim${30~cm} in length. We also produced a special ``installation jig'' for holding the polyethylene sheath vertically and for ease of manipulation. Several systems are used for temperature control of various part to accomplish cryo-fits, including a double walled dewar filled with liquid nitrogen for cooling and a commercial sous vide immersion heater for warming. A diagram illustrating installation steps is shown in Fig.~\ref{fig:HVFTassy}.

Prior to assembly, all pieces were cleaned by wiping down with ethanol and allowed to dry. The insulating HDPE sheath is placed inside a plastic bag and then submerged in a water bath warmed to $65^{\circ}$C using the immersion heater. The plastic bag keeps water from entering the polyethylene inner cavity and impeding the installation, as trapped water will not compress and will inhibit the interference fit.  While the water is heating, the threaded rod is affixed to the wide end of the copper pin. After the insulator has come to temperature for 30~minutes the pin can be placed into the warmed and now expanded insulating tube from the short cone end as shown in Fig. \ref{fig:HVFTassy}. The assembly is then allowed to return to room temperature.

The grounding tube is placed securely on a base with room for the polyethylene to emerge from bottom end.  The installation jig is attached to the polyethylene/copper assembly and then placed in a dewar. The dewar is filled with liquid nitrogen until at least the widest section of the insulator is covered. This will enables safe and convenient handling with cryo-safety gloves. After at least 20 minutes of dwell time the polyethylene is removed and inserted into the grounding tube, with the cold end up. If the parts have been properly sized, the UHMW contracts sufficiently to slide snugly into place against an internal lip without hammering or forcing.

Once the assembly has recovered to room temperature and the polyethylene has expanded, the installation jig is removed and the hex rod is used to insert and screw brass banana plug adapters in on both sides of the copper pin. Vacuum and hydrostatic pressure testing can be performed at this point. The conical steel cap is fabricated with excess material on the mating flange. The final operation is to machine the thickness of the flange down so the conic section of the polyethylene meets the interior conic section of the cap to provide an electrical boundary and avoid creepage current forming along the polyethylene surface. Once attached to the grounding tube, this completes the grounding layer around the insulator.

\subsection{Installation and cabling}
\label{sec:cabling}

\begin{figure}[t!]
    \centering
    \includegraphics[width=\textwidth]{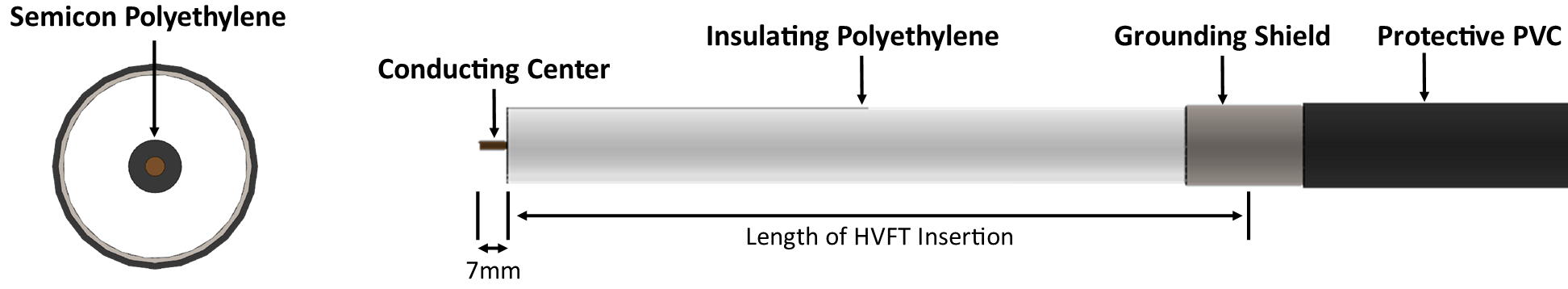}
    \caption{High voltage cable preparation for inserting into the HVFT from outside the vessel. Left shows a cross section view of the cable layers. Right shows the relative lengths each layer is stripped to before assembly.}
    \label{fig:cableprep}
\end{figure}

Installation of the HVFT onto the detector is a straightforward operation, though some cable preparation is required. The cables used on both sides of the feedthrough are Dielectric Sciences item 2134 which has 5 layers and is shown in Fig. \ref{fig:cableprep}. To begin, {7~mm} of the conductive center is exposed at the tips of the cables, removing the outer layers. Within the vessel, the cable is stripped entirely of its PVC jacket and grounding braid (labeled as "grounding shield" in Fig.~\ref{fig:cableprep}). Internally to the vessel there is no ground braid on the cable, which spirals up the field cage from low voltage toward the high voltage cathode, where the potential difference between the inside and outside of the cable becomes zero. The exposed conductive centers of both the internal and external cables connecting to the HVFT have banana plugs affixed to them via conductive epoxy, specifically, two-part MG Chemicals 8331-14G Silver Conductive Epoxy High Conductivity.

\begin{figure}[t]
    \centering
    \includegraphics[height=.44\textwidth]{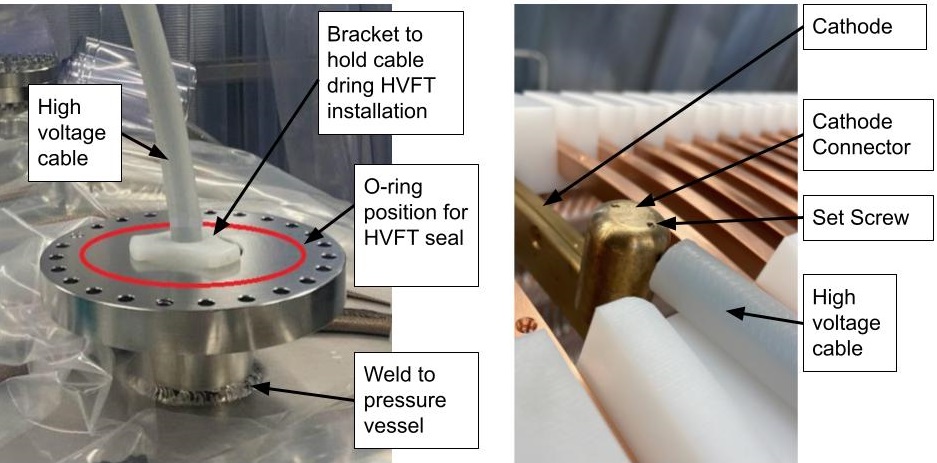}
    \caption{Left: UHMW clamp used to hold cable secure in the flange so it does not back off during installation into the HVFT. The red circle is a visualization of where the o-ring will sit once assembled. Right: connection of HV to the cathode.}
    \label{fig:HVFTinstall}
\end{figure}

To connect to the cathode, center conductor requires a robust mechanical and electrical connection on the cathode. For this purpose, we machined a custom connector piece from brass, as shown in Fig. \ref{fig:HVFTinstall} right. It has rounded and smoothed edges to minimize drastic electric fields arising from sharp points. The connector has a slot on the bottom that matches the width of the cathode edge and is secured in place on the cathode frame with brass set screws.  The conductive center of the cable inserts into a hole on the side of the connector and secures with additional brass set screws from the top. The cable is measured to be long enough to wrap helically around the field cage from the cathode to then emerge from a port at the top of the pressure vessel over the EL region. Enough cable is extended out from within the vessel to be fully inserted into the HVFT.

To install the HVFT onto the vessel, the assembled feedthrough is lifted above the port and then brought down close to the flange. The high voltage cable emerging from the detector can then be guided through the center of the insulating polyethylene. When the banana plug reaches the copper pin there is some resistance before slotting in. The same resistance is needed to disconnect the banana plug, ensuring a solid electrical connection.  Due to the geometry of the feedthrough port, cable strain tends to cause the cable to bend and slip backward into the vessel before fully inserting. To prevent this, we use a cable clamp as shown in Fig. \ref{fig:HVFTinstall} Left. This is a custom-machined UHMW piece that clamps the cable to the feedthrough plastic set screws inserted on the sides. The diameter is larger than the port bore hole so that it cannot be pushed down into the vessel.

The HVFT mates to the vessel via a DN180 O-ring flange which provides both a mechanical fixture and a pressure seal. This flange is designed to take reusable rubber rings during prototyping stages or single-use ultra-high vacuum capable HELICOFLEX\textsuperscript{\textcopyright} rings  during operation. Either ring type is placed in a groove on the bottom of the HVFT, creating a seal as it is tightened against the flat surface seen in Fig. \ref{fig:HVFTinstall} Left. The bolts used to tighten the HVFT to the vessel provide an immediate electrical grounding of the outer feedthrough sheath to the vessel ground. The cable outside the vessel is stripped of its PVC layer the depth into the HVFT plus a few centimeters, while the grounding braid is a few centimeters shorter, shown in Fig. \ref{fig:cableprep} Right. The external cable is then inserted through the HVFT until the banana plug is seated in the copper pin. The exposed grounding braid on the cable is then spread over the conical top of the HVFT and secured with a metallic hose clamp providing a continuous grounding shield. The fully installed assembly is shown in Fig. \ref{fig:realpics} Right.

\section{External system components}
\label{sec:ext_sys}
The cathode voltage for NEXT-100 is supplied by a commercial high voltage DC power supply (HCP 140-100000 from Fug).  The cable used for the HVFT has dimensions different from the output cable of the chosen high voltage power supply, 19.3~mm vs ~9.4~mm outer diameters respectively, and as such these two cables needed to be spliced externally to the vessel. This is done within a grounded oil-filled vessel (``splice pot'') that was modeled off a filter pot design used by the LArIAT experiment \cite{lariat_2020}.

The splice pot is made from a 100~Qt aluminum stewpot and filled with transformer oil, providing a volume of insulating material for the electrical connection between the two cables to be made, suppressing possible electrical discharges and keeping the outer casing safely at ground. The lid is welded down, and a smaller opening is made in the top for an adapter flange with two KF fittings for incoming and outgoing cables. Attached to the KF fittings are two insulating tubes made from G10 that sit inside the transformer oil and are connected via a copper bar across the bottom.

\begin{figure}[t]
    \centering
    \includegraphics[width=.99\textwidth]{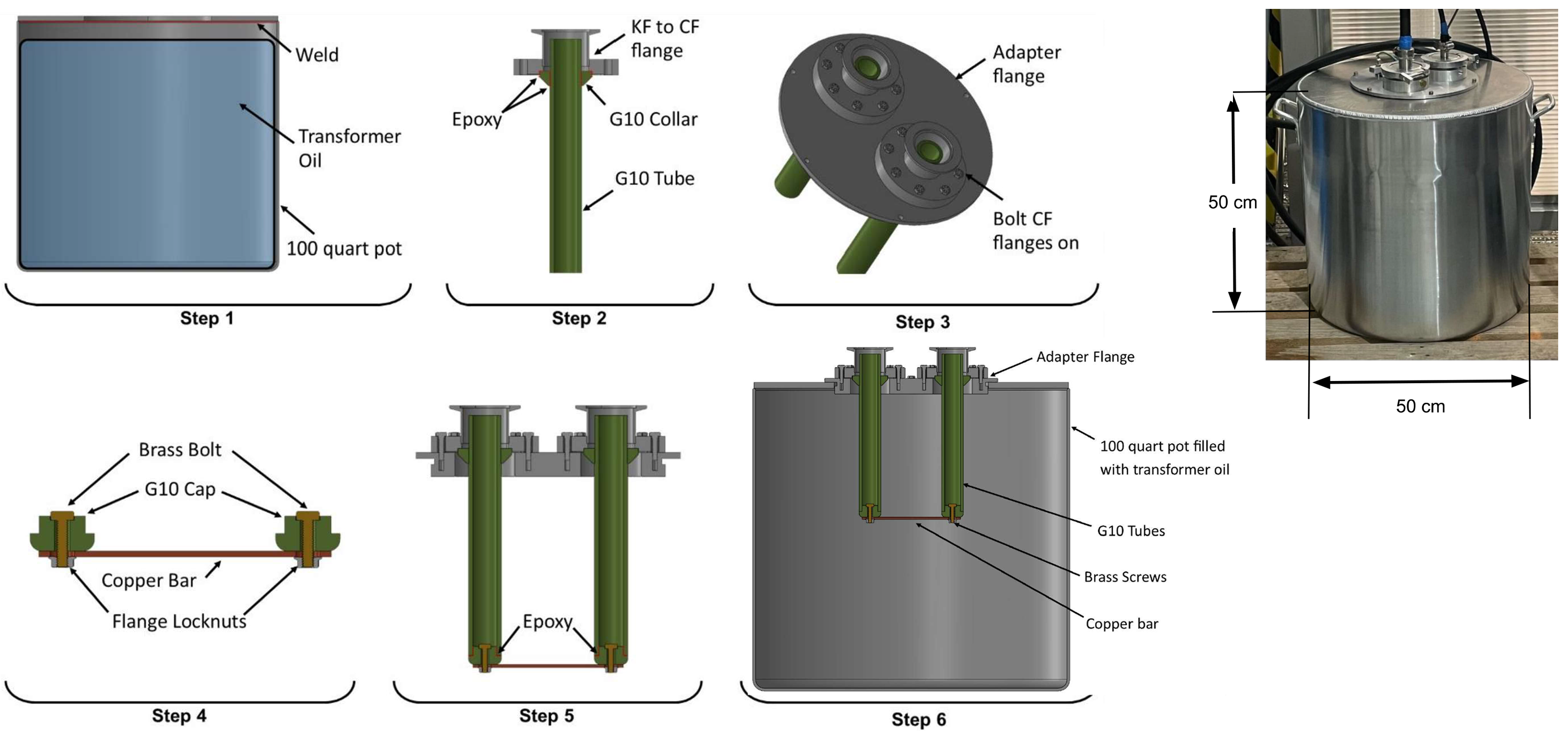}
    \caption{Assembly procedures for the Splice Pot with a picture of the NEXT-100 splice pot assembled and installed at the Laboratorio Subterráneo de Canfranc. Step 1 is filling the splice pot with transformer oil. Step 2 is building the G10 tube onto the KF flange followed by Step 3 where they are attached to the adapter flange. Step 4 is the copper bar assembly where the cables will make contact. Step 5 combines the G10 tubes to the copper bar assembly, and step 6 puts it into the transformer oil.}
    \label{fig:splicepot}
\end{figure}

To manufacture, a G10 collar was epoxied to a G10 tube and the KF flange. This was repeated twice and once the epoxy was fully cured, the KF flanges were be bolted onto the adapter flange. The copper bar was then attached to the ends of the G10 tubes using brass bolts and G10 caps, with the heads of the brass bolts creating the connections against the high voltage cables upon insertion. The adapter flange was then be bolted onto the pot with the copper assembly fully submerged in transformer oil. The pot is filled with enough room to insert the G10 tubes without spillages, and left filled for its operating lifetime. A diagram of the installation steps is shown in Fig. \ref{fig:splicepot}.

To ensure a solid electrical connection from the inner conductors of the cables to the brass screws at the bottom of the G10 tubes, a rounded brass thumbtack was pressed into the end of the conductive section of the cable.   The cables were stripped of their rubber casing a few cm longer than the depth of where the cable would reach from the top of the splice pot to the bottom of the G10 tubes, and the ground braid was stripped to {1-2~cm} shorter than that depth. The cable was then fed through its KF adapter before it is inserted into the top of the splice pot. Once the cable has reached the bottom, the KF adapter flange was secured to the pot, and the ground braid is splayed over the top. Pressing down on the cable to ensure a solid connection at the bottom, a hose clamp was then clamped down over the ground braid to hold the cable securely in place. To add a level of safety redundancy, the splice pot is grounded directly to the system ground as well using a braided cable attached to one of the lid bolts.



\section{Radiopurity\label{sec:radiopurity}}


The field cage presents one of the most serious radiopurity challenges since it contains a large mass of material and sits close to the active volume and is not shielded by the inner copper shield like the HVFT.  Careful selection and validation of field cage materials were thus required in order to meet the radioactive background specifications. Sourcing and validating these materials represented an important aspect of the system design and manufacture. The main radioactive backgrounds in NEXT originate from gamma rays via $^{208}$Tl (2.615 MeV) and $^{214}$Bi (2.448 MeV)  which are produced in the natural radioactive chains from $^{232}$Th and $^{238}$U respectively \cite{radiogenic,_lvarez_2013}. To check the appropriateness of materials, in most cases we assay for $^{232}$Th and $^{238}$U and assume secular equilibrium to predict the gamma activity. The individual materials were radioassayed at Pacific Northwest National Laboratory and the Laboratorio Subterráneo de Canfranc using an inductively-coupled plasma-mass spectrometer (ICP-MS) method. The measurements are shown in Table \ref{table:radioassayFC}.

\begin{table}[h]
\begin{center}
\centering
\begin{tabular}{ccccc}
\hline
Component  &  Material  &   $^{232}$Th [ppt] &      $^{238}$U [ppt] & Total Mass\\
\hline
Field cage resistors & - & $(0.22\pm 0.02)\times 10^6$ & $(0.28\pm 0.01)\times 10^6$ & 50 mg \\
Light Reflectors & PTFE & $2.6\pm 0.1$ & $4.9\pm 1.0 $ & 55.2 kg \\
Poly Wrap plus struts & HDPE & $15\pm 5$ & $3\pm 1$ &  88.3 kg \\
Field cage rings & Copper & <0.8 & <0.6 & 177.9 kg \\
\hline
\end{tabular}
\end{center}
\caption{The radioactivity measurements (arranged in decreasing total activity) and total mass of the components in the field cage. The resistors are made of several materials and are thus left unspecified. Only components that contribute to a total activity of $^{232}$Th and $^{238}$U greater than than 1~mBq are included.} 
\label{table:radioassayFC}
\end{table}

The total expected contribution to the background index of NEXT-100 from both the field cage and the HVFT will be acceptably low for both $^{208}$Tl and $^{214}$Bi contributions based on these results.

\section{Validation, commissioning and performance}
\label{sec:validation}

To validate the HVFTs, several tests were carried out including high-pressure validation, helium leak testing, high voltage performance under realistic detector conditions, and current and voltage stability measurements. 

\begin{figure}[t!]
    \centering
    \includegraphics[height=.5\textwidth]{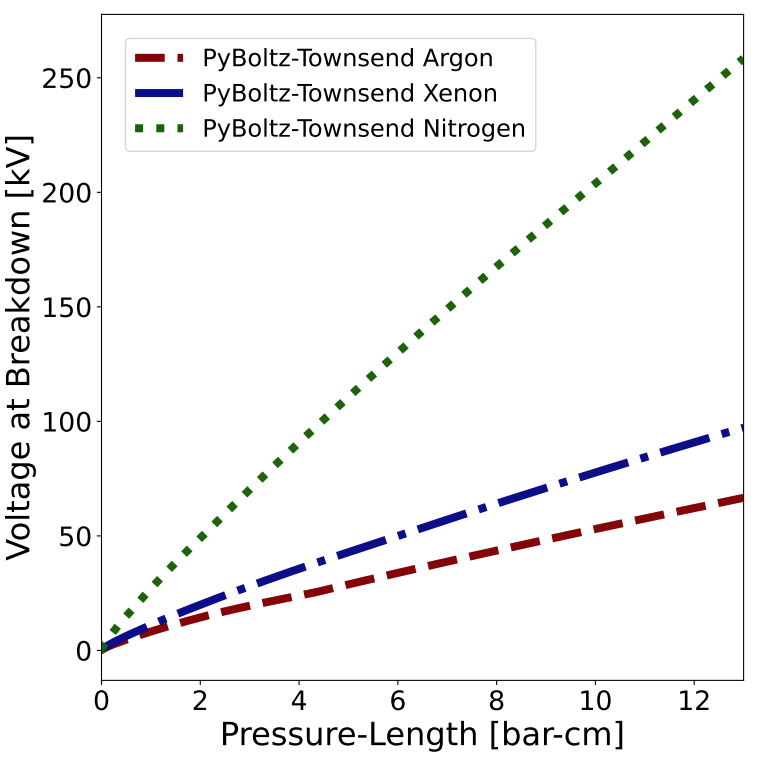}
    \caption{Breakdown voltage for various gases and pressures the HVFT are used with, simulated with PyBoltz \cite{ALATOUM2020107357,Norman_2022} using the Townsend model. }
    \label{fig:breakdown}
\end{figure}

To validate the high-pressure performance, the vessel attachment end was capped and the conical top removed so that the end could be attached to a hydrostatic testing machine. It was hydrostatically tested to 20.7~bar where it maintained that pressure for 10~minutes before being removed and was deemed safe for installation on a 15~bar max vessel. Each HVFT was then helium leak tested and demonstrated to be capable of vacuum lower than $10^{-5}$ ~Torr, with expectations of being significantly better than if put under vacuum alone rather than with a vessel filled with other components that are outgassing. Finally, each HVFT was installed onto the NEXT-100 prototype located at Argonne National Lab, called NEXT-CRAB, a scaled-up version of the system described in Ref.~\cite{byrnes2023next}. This allowed each to be tested under similar conditions to NEXT-100, including attachment to a field cage made of copper rings connected by a resistor chain that generates a uniform drift field in the detector. In this test, the vessel was first evacuated before being filled with 4.0~bar of nitrogen. Nitrogen has a higher breakdown point than xenon as shown in Fig. \ref{fig:breakdown} and therefore can demonstrate the HV capabilities at a lower total pressure. All feedthroughs were slowly ramped down to -70~kV, -5kV more than the required operating voltage of NEXT-100 for 400 V/cm drift field, and maintained at that voltage for at least 2 hours. During that time, there were no breakdowns and the voltage remained constant. The ramping up of voltage was operated at a maximum of 10~V/s to allow the field cage time to set at the voltage. This procedure has resulted in no electrical breakdowns or instabilities in the feedthroughs. Tests to find the maximum operating voltage of the feedthroughs were not carried out due to risk of damage caused by the discharges. 

\begin{figure}[t]
    \centering
    \includegraphics[height=.4\textwidth]{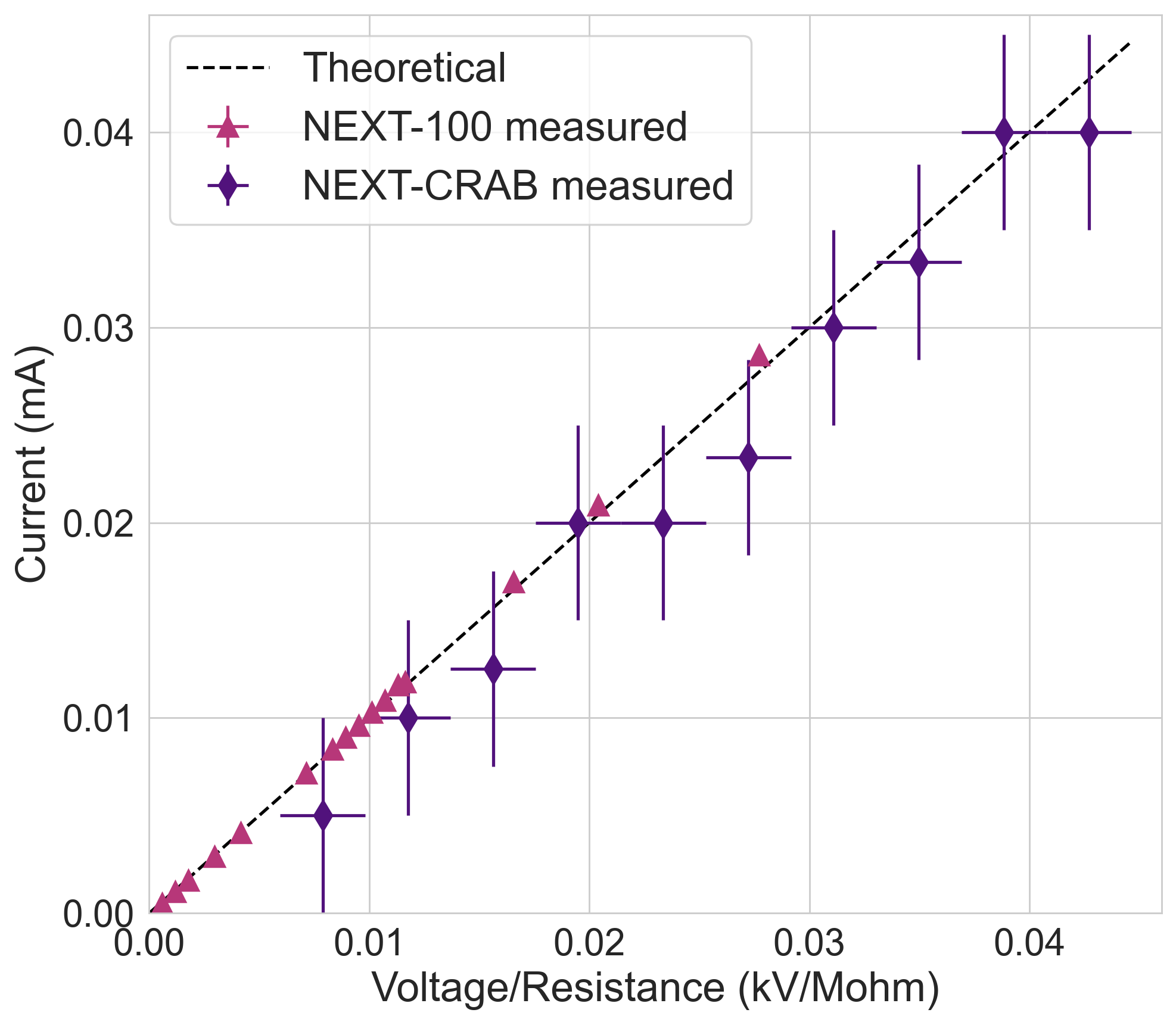}
    \includegraphics[height=.4\textwidth]{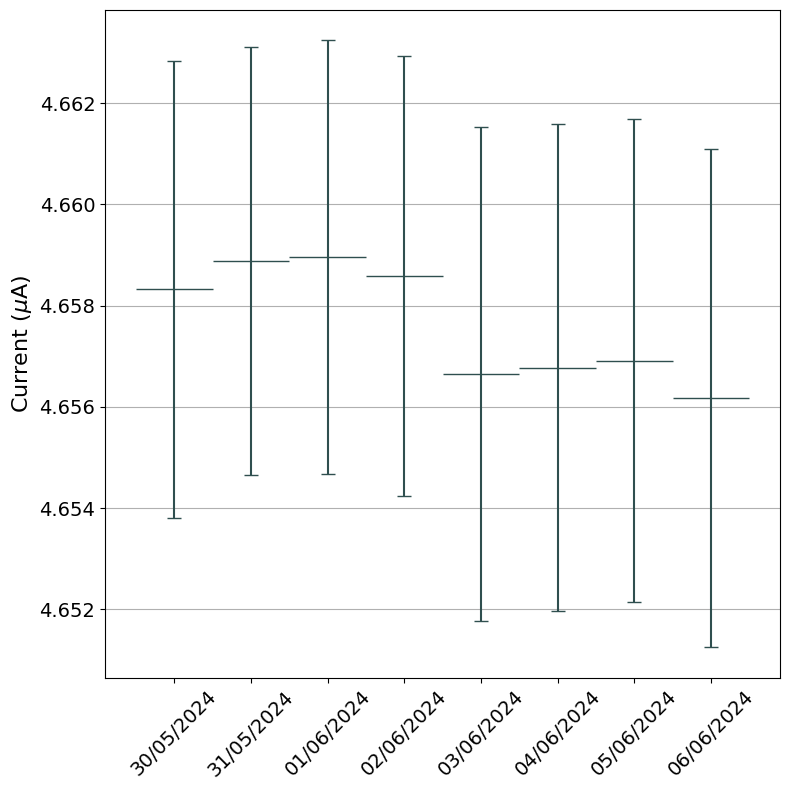}
    \caption{Left: Results from both HVFT and splice pot assemblies running in 4.0~bar of nitrogen gas for prototype NEXT-CRAB and for the NEXT-100 detector. The $y$ axis is the measured current and $x$ axis is calculated from the measured voltage and resistance across the field cages. Right: Current stability of NEXT-100 feedthrough in 4.2~bar of argon. }
    \label{fig:ohms}
\end{figure}

To validate the current and voltage stability and output, we compare the measured current at the power supplies with the predictions as calculated from Ohm's law. In these tests, the last ring of the field cage was also held at voltage to ensure a uniform and predictable drift field. The results of these measurements are shown in Fig.\ref{fig:ohms} left. The large error bars from NEXT-CRAB measurements are due to the low precision of the system's current readout. Also, because the resistors along the field cage are quite large ($\sim$50M$\Omega$), this results in relatively low current draws that never exceed 0.05~mA.

Following the validation of the feedthoughs in NEXT-CRAB, one HVFT remained on this system where it is now in use with 0.5~bar mixture of 2\% xenon and 98\% argon for continuing R\&D toward NEXT-HD~\cite{adams2021sensitivity}. The voltage has been demonstrated to reach and hold -70~kV consistently under these conditions over several months of intermittent operation and multiple ramps up and down. 

The other radiopure feedthough was installed and commissioned on the larger scale NEXT-100 vessel. During the commissioning, the detector was first operated in nitrogen gas before moving to an operation of several months in argon, and now currently in xenon gas for data taking at 4.2~bar. 
 The operation voltages on the cathode, pressures, and drift fields are shown in Table~\ref{tab:next100_performance}. The voltages in argon and xenon were chosen to operate below the expected breakdown based on the corresponding operation in nitrogen gas (Fig.~\ref{fig:breakdown}) and prior measurements and calculations~\cite{Norman_2022}. During these operations in different gas configurations, the voltage was stable with no electrical breakdowns. In addition, the cathode current has also been stable as shown in Fig.\ref{fig:ohms}, right. 

A higher-pressure run in xenon for NEXT-100 at 13.5~bar is expected once a reinforcement of the energy plane has been installed. Referring back to Fig. \ref{fig:breakdown}, the 4.2~bar measurements in xenon can be extrapolated to operation in xenon at 13.5~bar. This scaling yields $\sim$-69~kV, which is sufficient to meet the design specifications of a 400~V/cm drift field. 

These tests demonstrate the successful design, construction and validation of the NEXT-100 cathode HVFT design, which is successfully operating in the experiment. 

\begin{table}[h]
    \centering
    \begin{tabular}{l c c c}
        \hline
        Gas & Pressure [bar] & Voltage [kV] & Drift field [V/cm] \\
        \hline
        Nitrogen & 4.0 & -62.5 & 400 \\
        Argon & 4.2 & -14.7 & 65 \\
        Xenon & 4.2 & -23.0 & 146 \\
        \hline
    \end{tabular}
    \caption{Operational pressures, voltages at the cathode, and drift fields achieved during the commissioning and present operation of NEXT-100.}
    \label{tab:next100_performance}
\end{table}

\section{Conclusion}
\begin{figure}[t]
    \centering

    \includegraphics[height=.48\textwidth]{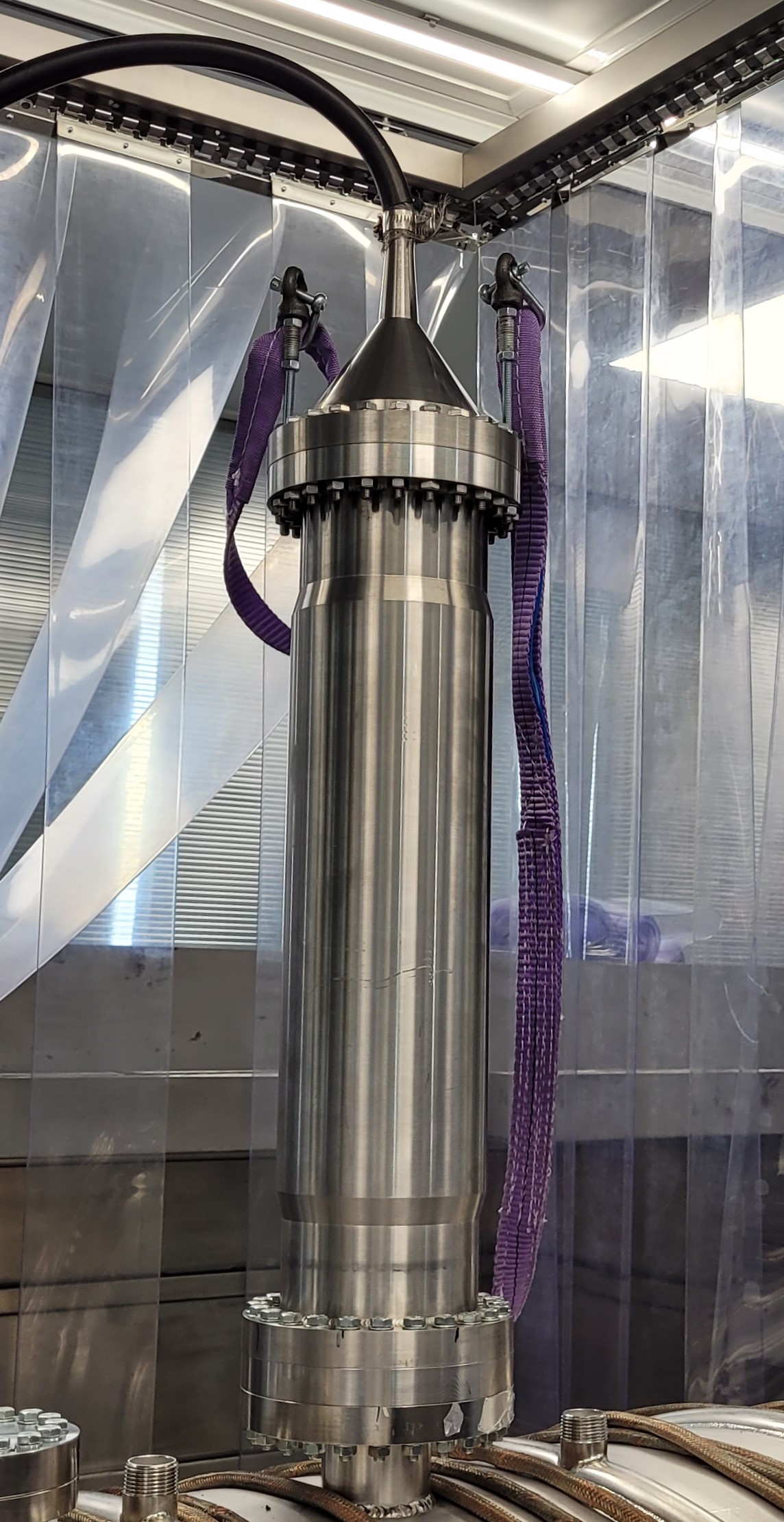}
    \includegraphics[height=.48\textwidth]{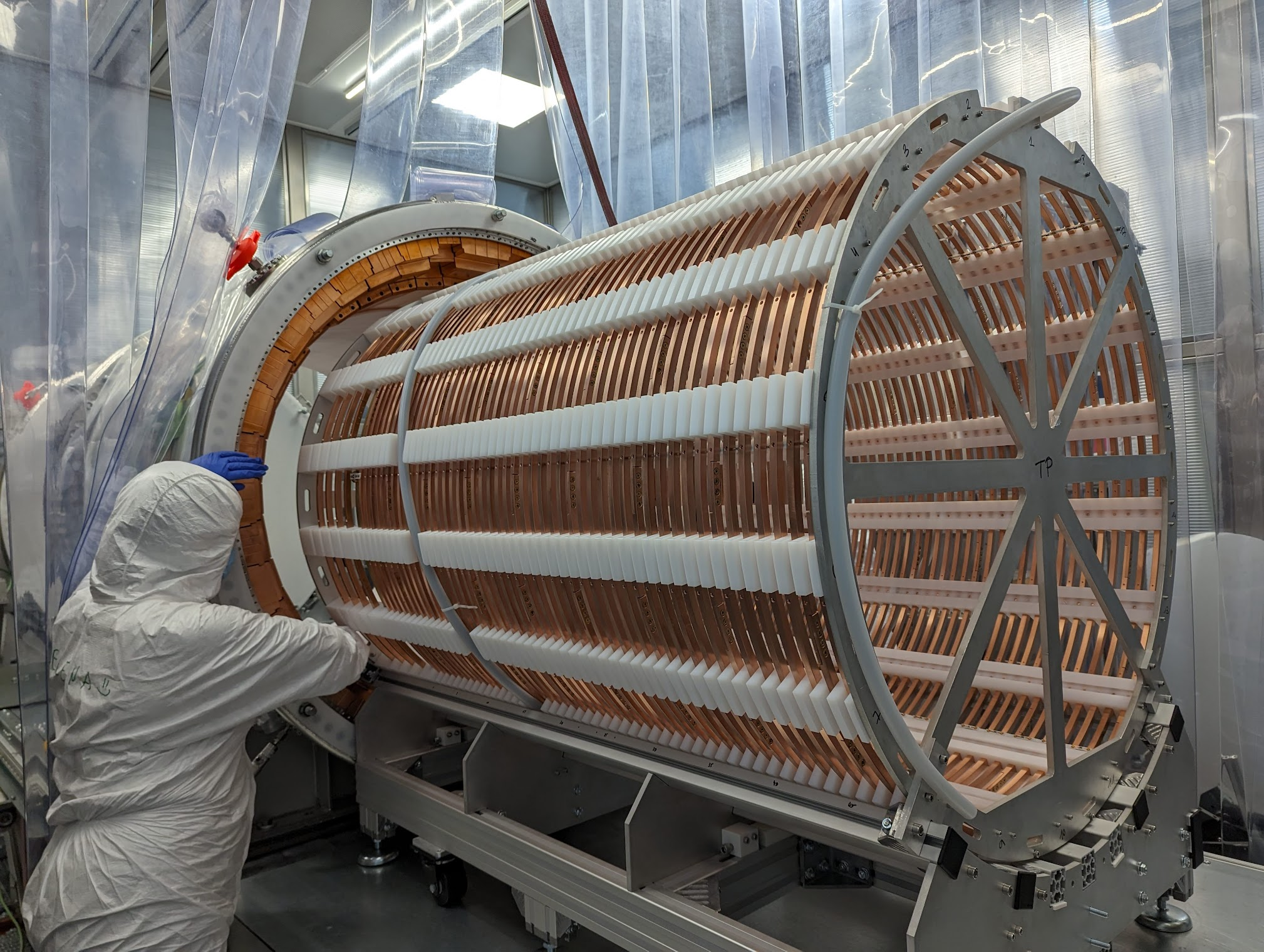}
    \caption{Left: HVFT installed on NEXT-100. Right: field cage prior to installation into the vessel.}
    \label{fig:realpics}
\end{figure}

The NEXT collaboration has demonstrated a robust cathode high voltage feedthrough (Fig.~\ref{fig:realpics}, left) design and distribution system that can stably provide voltages to at least -65~kV via a radiopure field cage (Fig.~\ref{fig:realpics}, right). The system was manufactured according to the principles of minimizing radioactivity and maximizing HV robustness.   System operation to date has been successful, achieving the design high voltage quickly during the commissioning period of NEXT-100 in various drift media.  The feedthrough design is adaptable to future TPCs and appears scalable for next-generation systems that require higher voltages and larger detector scales.  This system is operating now at the heart of the NEXT-100 experiment, which will proceed to search for $0\nu\beta\beta$ using high pressure xenon gas.

\section*{Acknowledgements}
The NEXT Collaboration acknowledges support from the following agencies and institutions: the European Research Council (ERC) under Grant Agreement No. 951281-BOLD; the European Union’s Framework Programme for Research and Innovation Horizon 2020 (2014–2020) under Grant Agreement No. 957202-HIDDEN; the MCIN/AEI of Spain and ERDF A way of making Europe under grants PID2021-125475NB and RTI2018-095979, and the Severo Ochoa and Mar\'ia de Maeztu Program grants CEX2023-001292-S, CEX2023-001318-M and CEX2018-000867-S; the Generalitat Valenciana of Spain under grants PROMETEO/2021/087 and CISEJI/2023/27; the Department of Education of the Basque Government of Spain under the predoctoral training program non-doctoral research personnel; the Spanish la Caixa Foundation (ID 100010434) under fellowship code LCF/BQ/PI22/11910019; the Portuguese FCT under project UID/FIS/04559/2020 to fund the activities of LIBPhys-UC; the Israel Science Foundation (ISF) under grant 1223/21; the Pazy Foundation (Israel) under grants 310/22, 315/19 and 465; the US Department of Energy under contracts number DE-AC02-06CH11357 (Argonne National Laboratory), DE-AC02-07CH11359 (Fermi National Accelerator Laboratory), DE-FG02-13ER42020 (Texas A\&M), DE-SC0019054 (Texas Arlington) and DE-SC0019223 and DE-SC0024438 (Texas Arlington); the US National Science Foundation under award number NSF CHE 2004111; the Robert A Welch Foundation under award number Y-2031-20200401. Finally, we are grateful to the Laboratorio Subterr\'aneo de Canfranc for hosting and supporting the NEXT experiment.

\bibliographystyle{JHEP}
\bibliography{Bibliography}

\providecommand{\href}[2]{#2}\begingroup\raggedright\begin{thebibliography}{10}

\bibitem{alvarez2013near}
V.~{\'A}lvarez, F.~Borges, S.~C{\'a}rcel, J.~Castel, S.~Cebri{\'a}n, A.~Cervera, C.~A. Conde, T.~Dafni, T.~Dias, J.~D{\'\i}az, et~al., {\it Near-intrinsic energy resolution for 30--662 kev gamma rays in a high pressure xenon electroluminescent tpc},  {\em Nuclear Instruments and Methods in Physics Research Section A: Accelerators, Spectrometers, Detectors and Associated Equipment} {\bf 708} (2013) 101--114.

\bibitem{Alvarez2013DEMO}
V.~Álvarez et~al., {\it Initial results of next-demo, a large-scale prototype of the next-100 experiment},  {\em Journal of Instrumentation} {\bf 8} (2013), no.~04 P04002.

\bibitem{WHITE_TDR}
F.~Monrabal et~al., {\it The next white (new) detector},  {\em Journal of Instrumentation} {\bf 13} (dec, 2018) P12010.

\bibitem{NEXT:2019qbo}
{\bf NEXT} Collaboration, J.~Renner et~al., {\it {Energy calibration of the {NEXT}-White detector with 1\% resolution near Q$_{\beta\beta}$ of $^{136}$Xe}},  {\em J. High Energ. Phys.} {\bf 10} (2019) 230, [\href{http://arxiv.org/abs/1905.13110}{{\tt arXiv:1905.13110}}].

\bibitem{NEXT:2021dqj}
{\bf NEXT} Collaboration, P.~Novella et~al., {\it {Measurement of the $^{136}$Xe two-neutrino double beta decay half-life via direct background subtraction in {NEXT}}},  {\em Phys. Rev. C} {\bf 105} (May, 2022) [\href{http://arxiv.org/abs/2111.11091}{{\tt arXiv:2111.11091}}].

\bibitem{NEXT:2023jsn}
{\bf NEXT} Collaboration, {C. Adams and others}, {\it Demonstration of neutrinoless double beta decay searches in gaseous xenon with next},  \href{http://arxiv.org/abs/2305.09435}{{\tt arXiv:2305.09435}}.

\bibitem{Alvarez2012NEXT100TDR}
V.~Álvarez et~al., {\it Next-100 technical design report (tdr). executive summary},  {\em JINST} {\bf 7} (2012), no.~06 T06001.

\bibitem{universe5030073}
V.~A. Kudryavtsev, {\it Recent results from lux and prospects for dark matter searches with lz},  {\em Universe} {\bf 5} (2019), no.~3.

\bibitem{t2knd280tpccollaboration2010time}
{T2K ND280 TPC collaboration}, {\it {Time Projection Chambers for the T2K Near Detectors}},  2010.

\bibitem{ACKERMANN2010141}
K.~Ackermann et~al., {\it A study with a small prototype tpc for the international linear collider experiment},  {\em Nuclear Instruments and Methods in Physics Research Section A: Accelerators, Spectrometers, Detectors and Associated Equipment} {\bf 623} (2010), no.~1 141--143. 1st International Conference on Technology and Instrumentation in Particle Physics.

\bibitem{She_2023}
X.~She et~al., {\it Development of time projection chamber prototype integrated with uv laser tracks for the future circular e+e- collider},  {\em Journal of Instrumentation} {\bf 18} (jul, 2023) C07015.

\bibitem{Ferrario:2015kta}
{\bf NEXT} Collaboration, P.~Ferrario et~al., {\it {First proof of topological signature in the high pressure xenon gas TPC with electroluminescence amplification for the NEXT experiment}},  {\em JHEP} {\bf 01} (2016) 104, [\href{http://arxiv.org/abs/1507.05902}{{\tt arXiv:1507.05902}}].

\bibitem{TomShuttKilotonne}
T.~Shutt, ``Some design considerations for xlzd and beyond.'' \url{https://indico.slac.stanford.edu/event/8015/contributions/7088/attachments/3477/9650/KTonneXLZD.pdf}, 2023.
\newblock Accessed: 2025-04-14.

\bibitem{martinez2018calibration}
{\bf NEXT} Collaboration, G.~Mart{\'\i}nez-Lema, J.~H. Morata, B.~Palmeiro, A.~Botas, P.~Ferrario, F.~Monrabal, A.~Laing, J.~Renner, A.~Sim{\'o}n, A.~Para, et~al., {\it {Calibration of the NEXT-White detector using $^{83m}$Kr decays}},  {\em JINST} {\bf 13} (2018), no.~10 P10014, [\href{http://arxiv.org/abs/1804.01780}{{\tt arXiv:1804.01780}}].

\bibitem{renner2019energy}
J.~Renner, G.~D{\'\i}az~L{\'o}pez, P.~Ferrario, J.~Hernando~Morata, M.~Kekic, G.~Mart{\'\i}nez-Lema, F.~Monrabal, J.~J. G{\'o}mez-Cadenas, C.~Adams, V.~{\'A}lvarez, et~al., {\it Energy calibration of the next-white detector with 1\% resolution near q$\beta$$\beta$ of 136xe},  {\em Journal of High Energy Physics} {\bf 2019} (2019), no.~10 1--13.

\bibitem{next100CDR}
{\bf NEXT} Collaboration, {Álvarez, V. and others}, {\it The {NEXT}-100 experiment for neutrinoless double beta decay searches ({Conceptual Design Report})},  \href{http://arxiv.org/abs/1106.3630}{{\tt arXiv:1106.3630}}.

\bibitem{Mistry_2024}
K.~Mistry et~al., {\it Design, characterization and installation of the next-100 cathode and electroluminescence regions},  {\em Journal of Instrumentation} {\bf 19} (Feb., 2024) P02007.

\bibitem{alvarez2013radiopurity}
V.~Alvarez, I.~Bandac, A.~Bettini, F.~Borges, S.~Carcel, J.~Castel, S.~Cebri{\'a}n, A.~Cervera, C.~Conde, T.~Dafni, et~al., {\it Radiopurity control in the next-100 double beta decay experiment: procedures and initial measurements},  {\em Journal of Instrumentation} {\bf 8} (2013), no.~01 T01002.

\bibitem{haefner2023reflectance}
J.~Haefner, A.~Fahs, J.~Ho, C.~Stanford, R.~Guenette, C.~Adams, H.~Almaz{\'a}n, V.~{\'A}lvarez, B.~Aparicio, A.~Aranburu, et~al., {\it Reflectance and fluorescence characteristics of ptfe coated with tpb at visible, uv, and vuv as a function of thickness},  {\em Journal of Instrumentation} {\bf 18} (2023), no.~03 P03016.

\bibitem{neves2017measurement}
F.~Neves, A.~Lindote, A.~Morozov, V.~Solovov, C.~Silva, P.~Bras, J.~Rodrigues, and M.~Lopes, {\it Measurement of the absolute reflectance of polytetrafluoroethylene (ptfe) immersed in liquid xenon},  {\em Journal of Instrumentation} {\bf 12} (2017), no.~01 P01017.

\bibitem{haefner2017reflectance}
J.~Haefner, A.~Neff, M.~Arthurs, E.~Batista, D.~Morton, M.~Okunawo, K.~Pushkin, A.~Sander, S.~Stephenson, Y.~Wang, et~al., {\it Reflectance dependence of polytetrafluoroethylene on thickness for xenon scintillation light},  {\em Nuclear Instruments and Methods in Physics Research Section A: Accelerators, Spectrometers, Detectors and Associated Equipment} {\bf 856} (2017) 86--91.

\bibitem{kravitz2020measurements}
S.~Kravitz, R.~Smith, L.~Hagaman, E.~Bernard, D.~McKinsey, L.~Rudd, L.~Tvrznikova, G.~O. Gann, and M.~Sakai, {\it Measurements of angle-resolved reflectivity of ptfe in liquid xenon with ibex},  {\em The European Physical Journal C} {\bf 80} (2020), no.~3 262.

\bibitem{nextcollaboration2025next100detector}
{NEXT Collaboration and C. Adams and others}, {\it {The NEXT-100 Detector}},  2025.

\bibitem{MechEngDesign}
R.~Budynas and J.~Nisbet, {\em Shigley's Mechanical Engineering Design, Tenth Edition}.
\newblock McGraw-Hill Education, 2015.

\bibitem{HVinsulators}
{\bf NEXT} Collaboration, L.~Rogers et~al., {\it High voltage insulation and gas absorption of polymers in high pressure argon and xenon gases},  {\em JINST} {\bf 13} (2018), no.~10 P10002, [\href{http://arxiv.org/abs/1804.04116}{{\tt arXiv:1804.04116}}].

\bibitem{plasticsmaterials}
J.~Brydson, {\em Plastics Materials, Seventh Edition}.
\newblock Butterworth-Heinemann, 1999.

\bibitem{MACUVELE20171248}
D.~L.~P. Macuvele, J.~Nones, J.~V. Matsinhe, M.~M. Lima, C.~Soares, M.~A. Fiori, and H.~G. Riella, {\it Advances in ultra high molecular weight polyethylene/hydroxyapatite composites for biomedical applications: A brief review},  {\em Materials Science and Engineering: C} {\bf 76} (2017) 1248--1262.

\bibitem{lariat_2020}
R.~Acciarri et~al., {\it The liquid argon in a testbeam (lariat) experiment},  {\em Journal of Instrumentation} {\bf 15} (apr, 2020) P04026.

\bibitem{radiogenic}
P.~Novella et~al., {\it Radiogenic backgrounds in the next double beta decay experiment},  {\em Journal of High Energy Physics} {\bf 2019} (oct, 2019).

\bibitem{_lvarez_2013}
V.~Álvarez et~al., {\it Radiopurity control in the next-100 double beta decay experiment: procedures and initial measurements},  {\em Journal of Instrumentation} {\bf 8} (Jan., 2013) T01002–T01002.

\bibitem{byrnes2023next}
N.~Byrnes, I.~Parmaksiz, C.~Adams, J.~Asaadi, J.~Baeza-Rubio, K.~Bailey, E.~Church, D.~Gonz{\'a}lez-D{\'\i}az, A.~Higley, B.~Jones, et~al., {\it Next-crab-0: a high pressure gaseous xenon time projection chamber with a direct vuv camera based readout},  {\em Journal of Instrumentation} {\bf 18} (2023), no.~08 P08006.

\bibitem{ALATOUM2020107357}
B.~{Al Atoum}, S.~Biagi, D.~González-Díaz, B.~Jones, and A.~McDonald, {\it Electron transport in gaseous detectors with a python-based monte carlo simulation code},  {\em Computer Physics Communications} {\bf 254} (2020) 107357.

\bibitem{Norman_2022}
L.~Norman, K.~Silva, B.~Jones, et~al., {\it Dielectric strength of noble and quenched gases for high pressure time projection chambers.},  {\em Eur. Phys. J. C} {\bf 82} (Jan, 2022) [\href{http://arxiv.org/abs/2107.07521}{{\tt arXiv:2107.07521}}].

\bibitem{adams2021sensitivity}
C.~Adams, V.~{\'A}lvarez, L.~Arazi, I.~Arnquist, C.~Azevedo, K.~Bailey, F.~Ballester, J.~Benlloch-Rodr{\'\i}guez, F.~I. Borges, N.~Byrnes, et~al., {\it Sensitivity of a tonne-scale next detector for neutrinoless double-beta decay searches},  {\em Journal of High Energy Physics} {\bf 2021} (2021), no.~8 1--24.

\end{thebibliography}\endgroup


\end{document}